\documentclass[aps,prb,twocolumn,groupedaddress,showpacs,showkeys,floatfix]{revtex4}
\usepackage{amsmath}
\usepackage{amssymb}
\usepackage{graphicx}
\usepackage{fixmath}
\usepackage[hang,nooneline,raggedright]{subfigure}

\newcommand*{\Exp}[1]{\ensuremath{\mathrm e^{#1}}}
\newcommand*{\dif}{\ensuremath{\mathrm d}}
\newcommand*{\dx}{\ensuremath{\dif x}}

\newcommand*{\dt}{\ensuremath{\dif t}}

\newcommand*{\I}{\ensuremath{\mathrm i}}

\newcommand*{\vect}[1]{\ensuremath{\mathbf{#1}}}

\DeclareMathOperator{\Dn}{dn}
\DeclareMathOperator{\Cn}{cn}
\DeclareMathOperator{\Sd}{sd}
\DeclareMathOperator{\Sn}{sn}

\DeclareMathOperator{\GAMMA}{\Gamma}

\begin{document}
\preprint{}
\title{Phason Dynamics in One-Dimensional Lattices}
\author{Hansj\"org Lipp}
\email{lipp@itap.physik.uni-stuttgart.de}
\author{Michael Engel}
\author{Steffen Sonntag}
\author{Hans-Rainer Trebin}
\affiliation{Institut f\"ur Theoretische und Angewandte Physik, University of
  Stuttgart, Pfaffenwaldring 57, 70550 Stuttgart, Germany}

\begin{abstract}
  In quasicrystals, the phason degree of freedom and
  the inherent anharmonic potentials lead to complex dynamics which cannot
  be described by the usual phonon modes of motion.
  We have constructed simple one-dimensional model systems, the dynamic Fibonacci chain (DFC)
  and approximants thereof.
  They allow us to
  study the dynamics of periodic and quasiperiodic structures with anharmonic double well
  interactions both by analytical calculations and by molecular dynamics simulations.
  We found soliton modes like breathers and kink solitons and
  we were able to obtain closed analytical solutions for special cases, which are in good
  agreement with our simulations.
  Calculation of the specific heat shows an increase above the Dulong-Petit value, which is due
  to the anharmonicity of the potential and not caused by the phason degree of freedom.
\end{abstract}


\maketitle

\section{Introduction}\label{sec:intro}

Quasicrystals\cite{qgbeug,lefsth} are aperiodic crystals with incommensurate
spatial frequencies due to noncrystallographic symmetries.
Their structure can be modelled by the projection of a planar, narrow stripe
(acceptance stripe) of a higher-dimensional crystal onto physical space\cite{janssen}.
As a consequence, the number of basis vectors in reciprocal space is higher
than the dimension of physical space, which leads to the possibility of new
symmetries not allowed in periodic crystals like e.g.\ five-fold or
icosahedral symmetry.

The dynamics of quasicrystals is governed by a complicated potential energy
landscape with more minima than atoms. Most of the time the atoms stay in
their respective local minima. However, on a picosecond time scale\cite{cod},
they can
overcome the energy barrier and swap into a
neighboring minimum.  To stress the instantaneous nature of this process, it
is called a flip.
The occurrence of these flips is a characteristic feature of quasicrystals and
follows from the construction method in higher-dimensional space: If the
acceptance stripe is translated perpendicular to the physical space, then some
lattice points leave the stripe while others enter
it.  The result in
physical space are discrete jumps of atoms. In the continuum picture, such
fluctuations of the stripe are identified with internal degrees of freedom,
the so-called phason modes.  Together with the conventional phonon modes, they
make up the dynamics of quasicrystals.

Although phason modes have great influence on macroscopic physical properties
of quasicrystals, e.g. diffusion\cite{papdiff1,papdiff2},
elasticity\cite{papelast}, plasticity\cite{papdef1,papdef2,phasonwall},
fracture\cite{qkcrack}, and phase
transitions\cite{papphase}, there is little knowledge about the precise atomic
motion underlying the phason flips.  Nevertheless,
phason fluctuations have been observed via speckle patterns\cite{phasonfluct}
and the tails of Bragg peaks in the
structure factors\cite{phasonmodes}.
Furthermore, direct observations have been made
by resolving the structural rearrangements of large
atom clusters with high-resolution transmission electron microscopy
\cite{paptem}.
On the theoretical side, there are so far only stochastic and hydrodynamic
models\cite{paphyd}, but few atomistic approaches.

To study phason modes closer, we have recently introduced a
simple one dimensional model system, the dynamic Fibonacci chain (DFC)\cite{dfcpaper}
and its approximants.
It consists of a chain of
particles with classical interaction
potentials that allow for phason flips.
Depending on the initial conditions, the chain can be either periodic or
quasiperiodic. In that paper\cite{dfcpaper}, we focussed on the influence of phason
flips on the dynamic and static structure factors in reciprocal space,
and we could show that they mainly caused
a broadening of the phonon peaks. The
reason is the fundamental difference of phason flips and phonon modes: Phonons
are periodic and extended in time and space, which yields clear signals in
reciprocal space. In contrast, phasons are stochastically distributed. They
can be observed only indirectly in reciprocal space by their interactions with
phonons. The interaction is both static and dynamic. It is static, because the
propagation of phonons is hindered by the disorder introduced due to previous
phason flips. It is dynamic, since during the flip of a particle, the
effective interaction
is highly
anharmonic and the harmonic approximation for phonon modes
is not applicable. Together, these effects decrease the lifetime of phonons
and broaden the characteristic lines.

In this paper we want to study the particle motion in real space. Can phason
flips propagate?  Are there soliton modes which other anharmonic systems show?
These systems, for example the well known Frenkel-Kontorova (FK) chain
or its continuous analogon, the sine-Gordon (SG) system,
basically show three modes of motions\cite{fk}:
a) Kinks, that are ``topological solitons'' and describe
a propagating topological defect in the chain (described as ``translatorische Eigenbewegung'' by Seeger\cite{eigenb});
b) breathers, which are localized oscillating nonlinear soliton modes (``oszillatorische Eigenbewegung'');
c) for small amplitudes,
one can linearize the equations of motion and obtains the well known phonon modes.
Indeed, we will show that upon strong excitations the dynamic Fibonacci chain displays soliton
modes in the form of breathers and kinks.

Furthermore, we calculate the specific heat of the DFC and observe that it rises beyond the Dulong-Petit law
at high temperatures, as observed by Edagawa et al. in icosahedral Al-Mn-Pd\cite{edagawaheat}.

In section \ref{sec:msys}, we introduce the model system we used for our studies.
In section \ref{sec:short}, short chains are studied both by analytical approaches and by molecualar dynamics simulations.
Section \ref{sec:ls} describes the soliton modes we found in the periodic chain with double well interaction.
In section \ref{sec:dfc}, we present our studies of solitary modes and of thermodynamic properties of the DFC.

\section{Model system}\label{sec:msys}
	\begin{figure}
		\centering
		\includegraphics[width=1.0\linewidth]{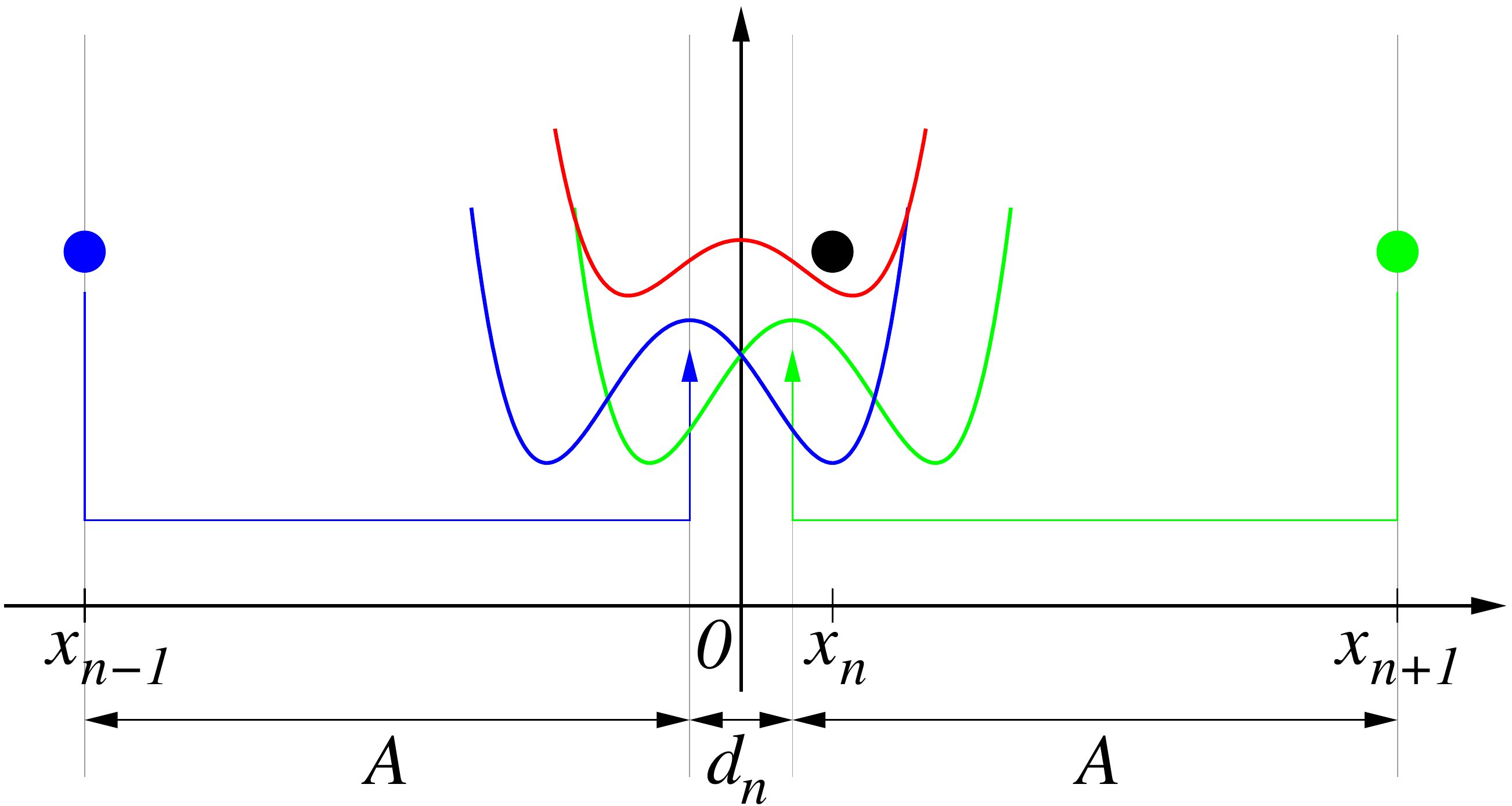}
		\caption{Superposition of the two double well potentials resulting from the pair interaction
		with the two neighbouring particles results in an effective potential of the form $a x^4+ bx^2$, where
		$b$ (and consequently the existence, position, and depth of energy minima) depends on the separation $d_n$
		of the centers of the underlying double well potentials.}
		\label{fig:dwdist}
	\end{figure}

The most simple quasiperiodic system is the one-dimensional Fibonacci chain.
It can be constructed by projecting a stripe of the two-dimensional square
lattice onto a straight line. 
There are two distinct distances of nearest neighbours in this chain, a short
one ($S$) and a long one ($L$). Their ratio equals the golden number
$\tau=\frac12(1+\sqrt5)\approx 1.618$. As the Fibonacci chain allows for flips
of the $LS$ pair to $SL$ and vice versa, we choose it as the
structure model for studying phason flips in quasicrystals.  Another structure
model we look at is the $LS$-chain, which is the periodic sequence of $L$ and
$S$. It is the simplest approximant of the Fibonacci chain with the
possibility of flips.

The interaction between neighboring particles is given by a model potential
that allows flips of atoms in $LS$ environments (two minima) and disallows
flips in $LL$ environments (one minimum) and $SS$ environments. Note
that the latter do not occur in the perfect Fibonacci chain, but might exist
in our systems.

It turns out that the best approach is an interaction potential of the
form\cite{dfcpaper}
\begin{equation}\label{eq:pot4}
  V_{\mathrm{part}}(x)=a x^4+ bx^2,
\end{equation}
where we usually use $a=1$ and $b=-2$.  In general, it is always possible to
transform an arbitrary potential of the form (\ref{eq:pot4}) to one with $a=1$
and $b\in\{-2,0,2\}$ by a coordinate change. $b<0$ is a double well with an energy barrier $E_0=\frac{b^2}{4a}$, and $b\ge 0$
has a single minimum.

The effective potential of particle $n$ is a superposition of the double
well potentials which left and right neighbour exert and is plotted in figure \ref{fig:dwdist}.
$A$ is the length of the double well ``spring'' and, for our standard choice $a=1, b=-2$, equals $2\tau+1$.
$d_n$=$x_{n+1}-x_{n-1}-2A$ is the separation of the centers of the double wells. If we choose the coordinate
origin in the middle of the neighbour particles, the potential reads:
\begin{align}
	\label{eq:neighbourpot}
	\begin{split}
V_{\mathrm{eff}}(x_n)&=V_{\mathrm{part}}\left(x_n-\frac{d_n}{2}\right)+V_{\mathrm{part}}\left(x_n+\frac{d_n}{2}\right)\\
	&= 2 \left(ax_n^4 + \tilde{b}x_n^2 + \Delta\right)
	\end{split}
\end{align}
	with $\tilde{b}=b + \frac{3}{2}ad_n^2$ and $\Delta=a(d_n/2)^4 + b(d_n/2)^2$.
	For $b<0$ (double well), there is a critical separation $d_\mathrm{crit}$, where
	the sign of
	$\tilde{b}$ changes:
	\begin{equation}
		\label{eq:dcrit}
		d_\mathrm{crit}=\sqrt{\frac23 \frac{|b|}{a}}
	\end{equation}
	So, there is a bifurcation: for $|d_n| < d_\mathrm{crit}$, we have a double well with minima at $x_{\pm}(d_n)=\pm\sqrt{\frac{|b|}{2a}-\frac34 d_n^2}$; for $|d_n|>d_\mathrm{crit}$, there is only one minimum at
	$x=0$.

	The total potential of the chain is
	\begin{equation}
		\label{eq:interact}
		V=\sum_n V_{\mathrm{part}}(x_{n+1}-x_n-A)
	\end{equation}
	For $x_{n+1}-x_{n-1}=2A= 2(2\tau+1)$ one obtains a double well in $LS$ environment with $L=2\tau+2$ and $S=2\tau$,
	for $x_{n+1}-x_{n-1}=2A\pm2$ one obtains single wells in $LL$ and $SS$ environments.

	In the two minima regime, the barrier height and distance of the minima are maximal, decreasing continuously when $d_n$ is approaching
	the critical distance.

\section{Short chains}\label{sec:short}

In our attempts to learn about analytically accessible solutions of the
equations of motion for chains underlying the above interaction potential we
started with short chains. If these are constructed with
periodic boundary conditions, they also describe periodic modes
of motion in infinite chains.

\begin{figure}
	\centering
	\includegraphics[width=\linewidth]{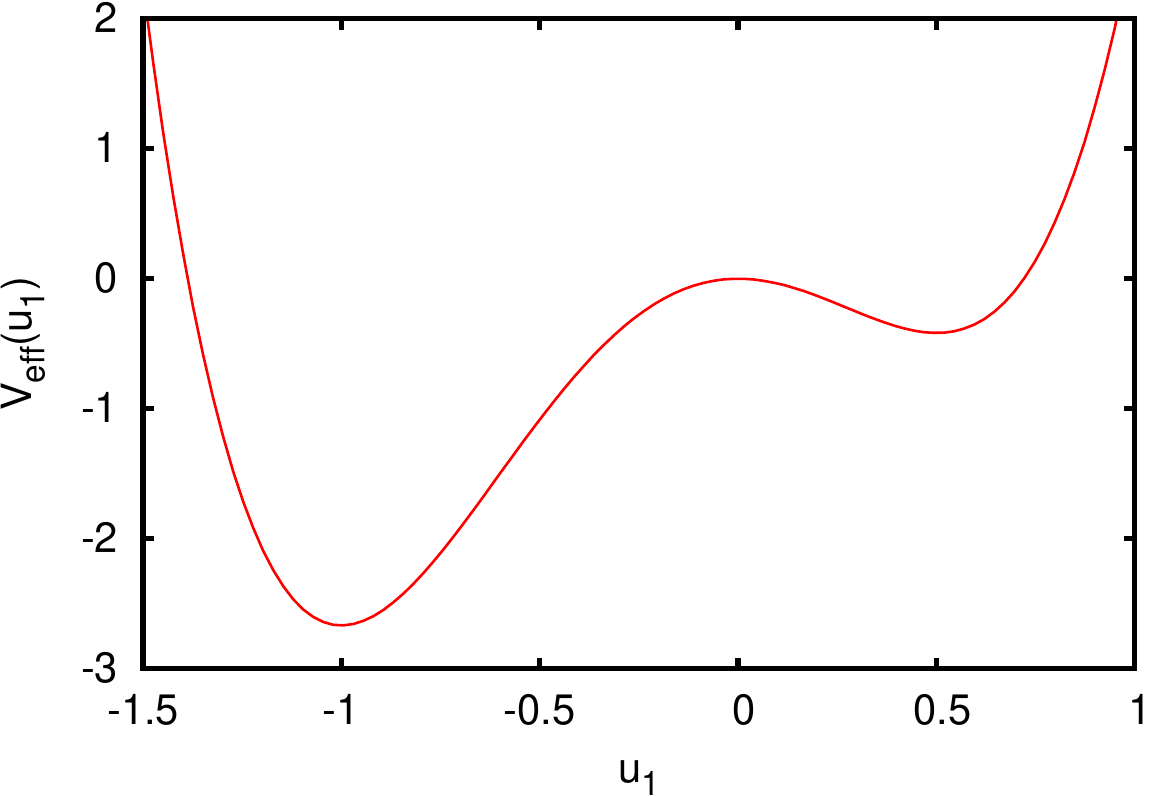}
	\caption{Effective potential $V_{\mathrm{eff}}(u_{1})$ a particle feels
	in the chain of three particles of length $LSL$ with the double-well interaction potential
	$V_{\mathrm{part}}(x)=x^4 - 2x^2$ if one particle is not moving ($D=1$, $a=1$, $b=-2$, $u_2=0$).
	As in previous work\cite{dfcpaper}, we use arbitrary units in this and the following
	figures.}
	\label{fig:ch3d1pot}
\end{figure}
\begin{figure}
	\centering
	\includegraphics[width=\linewidth]{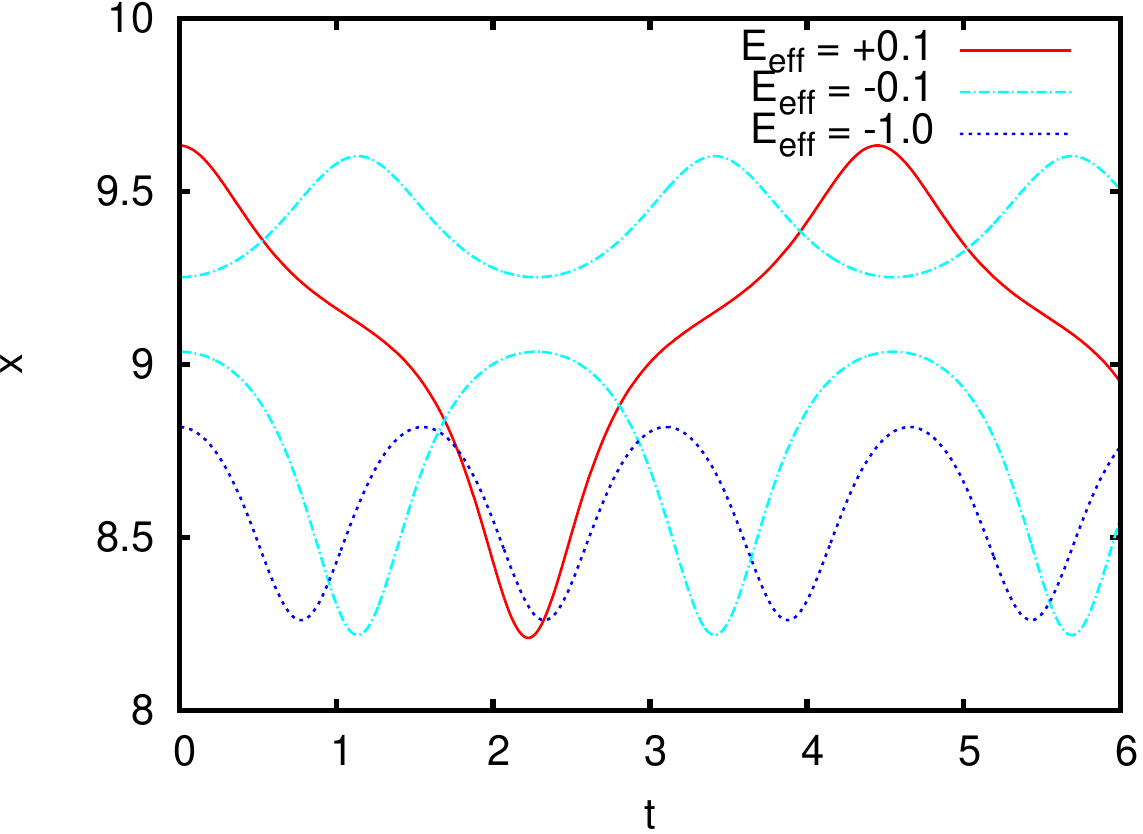}
	\caption{Trajectory $x_1(t)$ of the first particle of the chain of three particles of length $LSL$ with the double-well interaction potential
	$V_{\mathrm{part}}(x)=x^4 - 2x^2$ if one particle is not moving ($D=1$, $a=1$, $b=-2$, $u_2=0$, $M=1$) at three different effective energies.
	While there is only one trajectory for high ($E_\mathrm{eff}>0$) and low ($E_\mathrm{eff}<-\frac{5}{12}$) energies, there are two possible trajectories in the intermediate regime, because a particle
	can be situated  in one of the two energy minima shown in figure \ref{fig:ch3d1pot}.
	The trajectories of the simulated particles are not distinguishable from the analytical results.}
	\label{fig:ch3d1ana}
\end{figure}

Since the short ``chain'' of two particles can be described as a single particle
in a double well potential (or to be more precise: a potential of the form $V(x)=ax^4+bx^2$, $a>0$) with known solutions\cite{onod1}
we started with chains of three atoms and fixed length $L_\mathrm{C}=3A+D$, where $D$
describes the length difference relative to the chain where all particles have
the average distance $A$.

One approach to handle the hamiltonian
\begin{equation}
	\label{eq:ham}
	\begin{split}
		H(\vect{x},\vect{p}) &= \sum_{i=1}^3 \frac{p_i^2}{2M}
		+ V_{\mathrm{part}}(x_3-x_2-A) \\&\quad
		+ V_{\mathrm{part}}(x_1+L_\mathrm{C}-x_3-A) \\&\quad
		+ V_{\mathrm{part}}(x_2-x_1-A)
	\end{split}
\end{equation}
(later, we will use particle mass $M=1$)
is the coordinate transformation
\begin{equation}
	\begin{pmatrix} R \\ u_1 \\ u_2 \end{pmatrix}
	=
	\begin{pmatrix}
		\frac13 & \frac13 & \frac13 \\
		1 & - \frac12 & -\frac12 \\
		-\frac12 & 1 & -\frac12
	\end{pmatrix}
	\begin{pmatrix} x_1 \\ x_2 \\ x_3 \end{pmatrix}
	+
	\frac{L_\mathrm{C}}{2}\,
	\begin{pmatrix} 0 \\ 1 \\ 0 \end{pmatrix},
\end{equation}
where $R$ represents the coordinate of the center of mass. The transformation
leads to a coupled system of non-linear differential equations:
\begin{subequations}
	\label{eq:newtondgl}
	\begin{align}
		\ddot{u}_1 &:= - \frac{12}{M} \left(au_1^3+u_1\left(\frac{b}{2} + \frac13  a\left(2 u_2 + u_1 + D\right)^2\right)\right)\\
		\ddot{u}_2 &:= - \frac{12}{M} \left(au_2^3+u_2\left(\frac{b}{2} + \frac13  a\left(2 u_1 + u_2 - D\right)^2\right)\right)
	\end{align}
\end{subequations}
The only approach to decouple these equations for the system $a=1$, $b=-2$,
$D=1$ ($LSL$-Sequence which allows phason flips, because of the double well
potential) was setting $u_2=0$, i.e.\ one particle does not move at all.
The simplified equation of motion
\begin{equation}
	\ddot{u}_1 = -8\left(2u_1^3+ u_1^2 -u_1\right)
\end{equation}
then suggests an effective potential
\begin{equation}
	V_\mathrm{eff}(u_1) = 4\left(u_1^4+ \frac23u_1^3 -u_1^2\right)
\end{equation}
which is shown in figure \ref{fig:ch3d1pot}. The potential is not symmetric any more but has the zeroes $\frac{-1\pm\sqrt{10}}{3}$,
local minima $\left(-1;-\frac{8}{3}\right), \left(\frac12;-\frac{5}{12}\right)$, and the local maximum $(0;0)$.
The equation
$E_\mathrm{eff}=\frac12 M \dot{u}_1^2 + V_\mathrm{eff}(u_1)$ finally
leads to a trajectory
\begin{align}
	u_1 &= r_2 + \frac{r_2-r_1}
	{C\Sn^2\left((t-t_0)\Omega\Big|m\right)-1}\\
\intertext{with the roots $r_i$ of $E_\mathrm{eff}-V(u)=0$:}
	\prod_{i=1}^4 (u-r_i) &= u^4+ \frac23 u^3 - u^2 - \frac14 E_\mathrm{eff}\label{eq:root}\\
\intertext{and}
	C&=\frac{r_4-r_1}{r_4-r_2},\\
	m&=\frac{r_2-r_3}{r_1-r_3}\frac{r_4-r_1}{r_4-r_2},\\
	\Omega&=\sqrt{\frac{2(r_1-r_3)(r_4-r_2)}{M}}.
\end{align}
It is remarkable that the parameter $m$ of the Jacobi elliptic function
$\Sn$\cite{ww,abrsteg}
is not only energy dependent, it has also very untypical values: From the analysis of the
possible roots of equation (\ref{eq:root}) follows that the usual
condition $0\le m\le 1$ does not hold, but the admitted values for $m$ 
in the complex plane can have the following
forms:
\begin{subequations}
	\begin{align}
		m_1 &= \varphi \\
		m_2 &= \frac12 + \I\varphi \\
		m_3 &= \Exp{\I\varphi} \\
		m_4 &= 1 - \Exp{\I\varphi}
	\end{align}
\end{subequations}
with energy dependent $\varphi\in \mathbb R$.  So, $m$ is not necessarily a
real number between 0 and 1, but can lie on the straight line $m_2$ in the
complex plane, on the real axis, and on the circles $m_3$ and $m_4$. It is
worth noting, that the oscillation frequency $\Omega$ is energy dependent,
which is due to the anharmonicity of the potential.  Figure \ref{fig:ch3d1ana}
shows typical trajectories for different energies. There is only one
trajectory for high and low energies, but in the intermediate regime there are
two trajectories for the same energy, one for the particle motion in the deep
minimum of the effective potential (figure \ref{fig:ch3d1pot}) and one for the other minimum.

The observed behaviour is typical for the dynamics of quasicrystals. At lower energies, the particles oscillate
in their local minima. At intermediate energies, the particles can sit in different minima, but usually cannot overcome the
barriers inbetween.

\begin{figure}
	\centering
	\includegraphics[width=\linewidth]{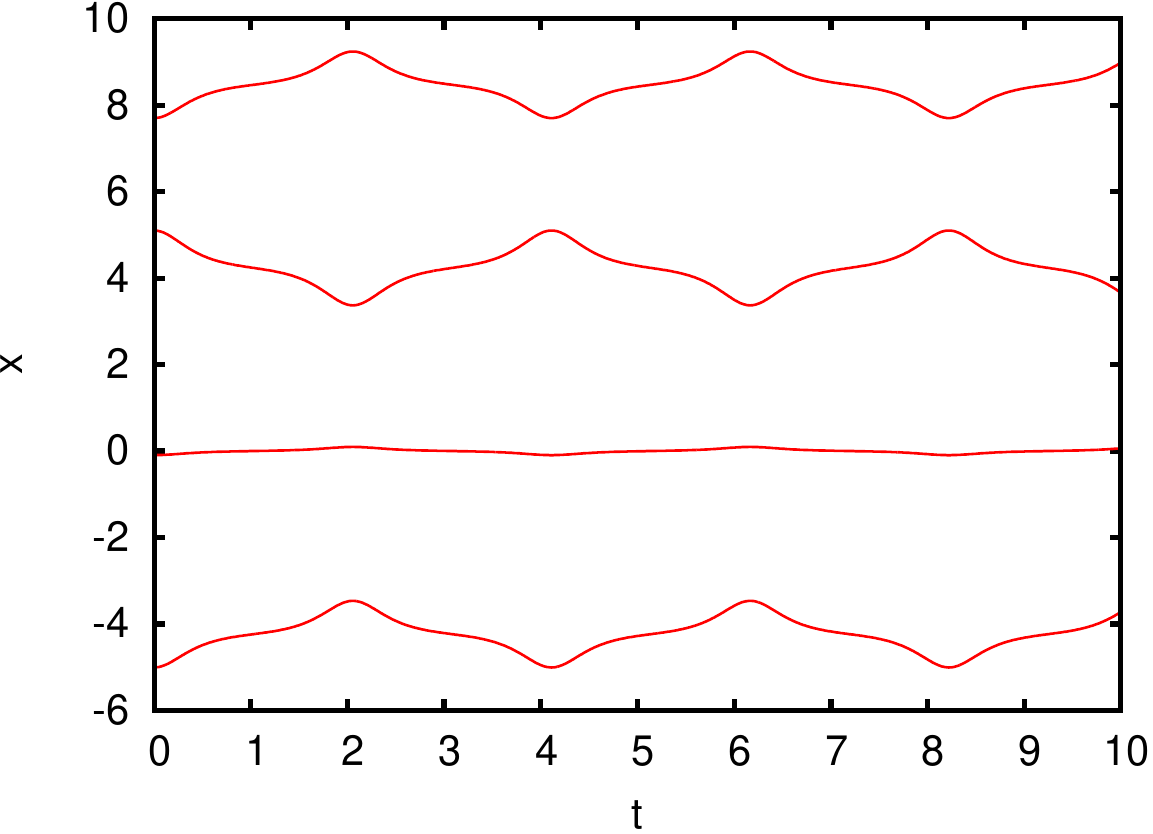}
	\caption{Trajectories of the particles in the chain of three particles of length $3A$ with the double-well interaction potential
	$V_{\mathrm{part}}(x)=x^4 - 2x^2$ ($D=0$, $a=1$, $b=-2$, $M=1$) with the polar angle $\varphi=0.1$ at $E=0.1$. Also in this case,
	the difference to the simulated trajectories is not visible.}
	\label{fig:ch3d0anasim05}
\end{figure}

As a more general approach to solve Eq.~(\ref{eq:ham}) for $D=0$ (the chain has the length $3A$), we used
Jacobi coordinates which are described in the classical solution\cite{kl} of
the three body problem \cite{Calogero}. After transforming to Jacobi
coordinates
\begin{equation}
	\begin{pmatrix} R \\ x \\ y \end{pmatrix}
	=
	\begin{pmatrix}
		\frac13 & \frac13 & \frac13 \\
		-\frac12 \sqrt2 & 0 & \frac12 \sqrt2\\
		\frac16 \sqrt6 & -\frac13 \sqrt6 & \frac16 \sqrt6
	\end{pmatrix}
	\begin{pmatrix} x_1 \\ x_2 \\ x_3 \end{pmatrix}
	-
	\frac{\sqrt2}{3}L_\mathrm{C}\,
	\begin{pmatrix} 0 \\ 1 \\ 0 \end{pmatrix}
\end{equation}
and then to polar coordinates, $x=r\cos\varphi$, $y=r\sin\varphi$, we
get a simpler Hamiltonian:
\begin{equation}
	H=\frac12 M \dot{r}^2+\frac12 Mr^2\dot{\varphi}^2+\frac32 M \dot{R}^2
	+ \frac92 a r^4 + 3br^2
\end{equation}
$R$ and $\varphi$ are cyclic variables. Thus, $\dot{R}$ and
$L_\varphi:=Mr^2\dot{\varphi}$ are constant. This leads to equation
\begin{equation}
	E=\frac12 M \dot{r}^2+  \frac{L_\varphi^2}{2M} r^{-2} + \frac92 a r^4 + 3br^2
\end{equation}
where $E$ already contains the motion of the center of mass.

So for $D=0$, the motion of three particles in one dimension can be treated as
the motion of one particle with generalized angular momentum $L_\varphi$ in
two dimensions.

For $a=1$ and $b=-2$ the solution is
\begin{align}
	\label{eq:solch3d0}
	r^2&=r_3+(r_2-r_3)\Sn^2\left(\Omega\cdot(t-t_0)\Big|m\right)\\
\intertext{with}
	\Omega&=3\sqrt{\frac{r_3-r_1}{M}}, \quad m=\frac{r_2-r_3}{r_1-r_3}
\end{align}
$r_i$ are the solutions of $\frac92 x^3 - 6x^2-Ex+\frac{L_\varphi^2}{2M}=0$:
\begin{equation}
	\prod_{i=1}^3 (x-r_i) = x^3 - \frac{4}{3}x^2-\frac29 Ex+\frac{L_\varphi^2}{9M}
\end{equation}

For $L_\varphi \ne 0$, we get the angle
\begin{equation}
	\varphi=\frac{L_\varphi}{M}\int\frac{\dt}{r_3+(r_2-r_3)\Sn^2\left(\Omega\cdot(t-t_0)\Big|m\right)}
\end{equation}
As there is no closed solution of this integral, only the case $L_\varphi=0$
has been solved. In this case, $\varphi$ is constant and equation (\ref{eq:solch3d0})
simplifies to
\begin{align}
	r &= \tilde{C}(E) \Cn\left(\tilde{\Omega}(E)(t-t_0)\,\Bigg|\,\tilde{m}(E)\right)\\
\intertext{with}
	\tilde{C}(E)&=\sqrt{\frac23+\frac13\sqrt{4+2E}}\\
	\tilde{\Omega}(E)&=\sqrt{\frac{6}{M}\sqrt{4+2E}}\\
	\tilde{m}(E)&=\frac12 + \frac{1}{\sqrt{4+2E}}
\end{align}
where $\Cn$ is another Jacobi elliptic function.

Here, the oscillation frequency is energy dependent, again. $r(t)$ resembles
the motion of one particle in a double well potential. The constant angle
$\varphi$ determines the ratios of $r(t)$ and the amplitudes $x_i(t)$ of the three
particles.
Figure \ref{fig:ch3d0anasim05} shows typical trajectories in this system. For
$E < 0$ the $\Cn$ function evolves to a $\Dn$ function, because of
$\tilde{m}>1$.

In this section, we obtained results for special modes of motion in our
systems showing interesting features like energy dependent oscillation
frequencies and disconnected trajectories of same energy in phase space.
Closed analytical solutions were found for a few special cases of particles oscillating in phase
which also describe collective modes of motions in large chains.

\section{Periodic $LS$ chain in real space}\label{sec:ls}

The next step towards the Fibonacci chain is the periodic $LS$ chain, a chain
of particles with alternating distances $L$ and $S$ and total
potential as in equation (\ref{eq:interact}) with $a=1$ and $b=-2$. The periodicity
admits analytical calculations. Further simulations suggest that
the results may be relevant for chains that contain defects like $SS$ or $LL$ sequences or for the
Fibonacci chain.

\subsection{Basic modes}

	\begin{figure}
		\subfigure[A breather decays in two kink solitons. Here, one particle of the periodic $LS$ chain
		was initialized with a start velocity $v(0)=-10$ (corresponding to a kinetic energy
		$50 \gg$ height 1 of the potential barrier),
		resulting in the unstable breather and two kink solitons.]{\label{fig:soltraj}%
		\includegraphics[width=\linewidth]{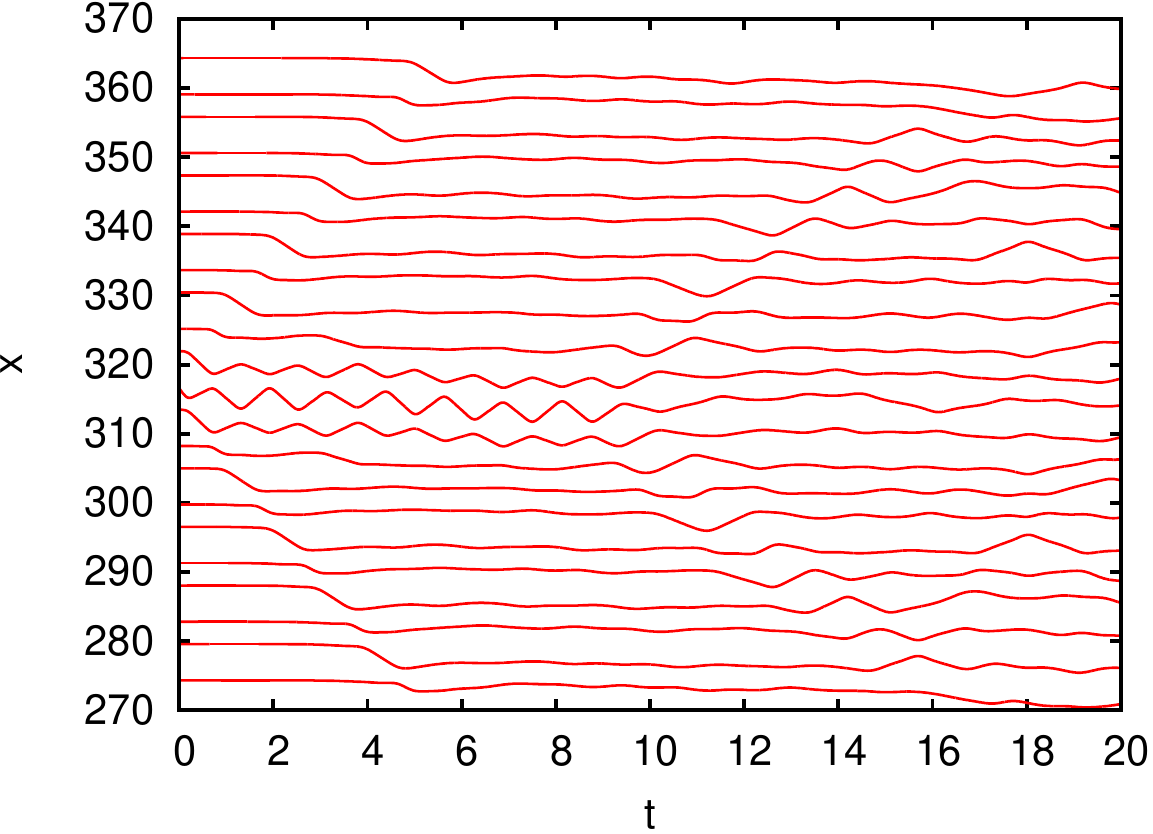}}%
		\\%
		\subfigure[Two kink solitons crossing with little interaction.
		]{\label{fig:kinkcol}%
		\includegraphics[width=\linewidth]{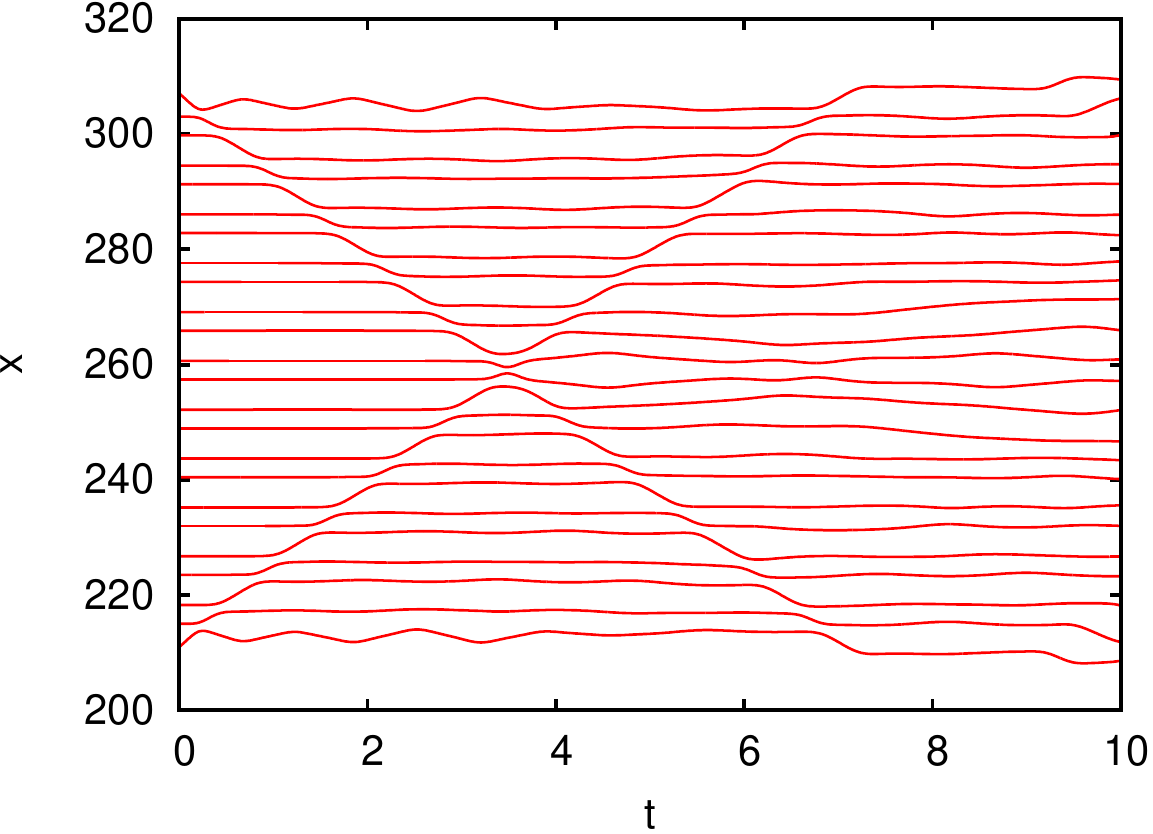}}
		\caption{Modes of motion in the periodic $LS$ chain}
	\end{figure}

We study this chain both by analytical approaches and by a numerical solution
of the equations of motion with molecular dynamics simulations. In both cases
periodic macro boundary conditions are applied.

Figure \ref{fig:soltraj} shows some exemplary trajectories: After providing one
particle with an initial velocity, we observe three solitary waves: One breather
and two kink solitons propagating to both sides of the chain. These modes indeed show
characteristics of the well-known modes of the sine-Gordon model. For example,
the solitons can cross each other without noticeable interaction (figure \ref{fig:kinkcol}).
Other initial conditions can lead to soliton-soliton interaction, which is
not present in the standard sine-Gordon model. In the figure, the breather
decays into two kink solitons.

Besides the kinks and breathers, naturally there are further modes:
The rippled curves at large $t$ are due to phonons in the system.

\subsection{Kinks}
	\begin{figure}
		\centering
		\includegraphics[width=1.0\linewidth]{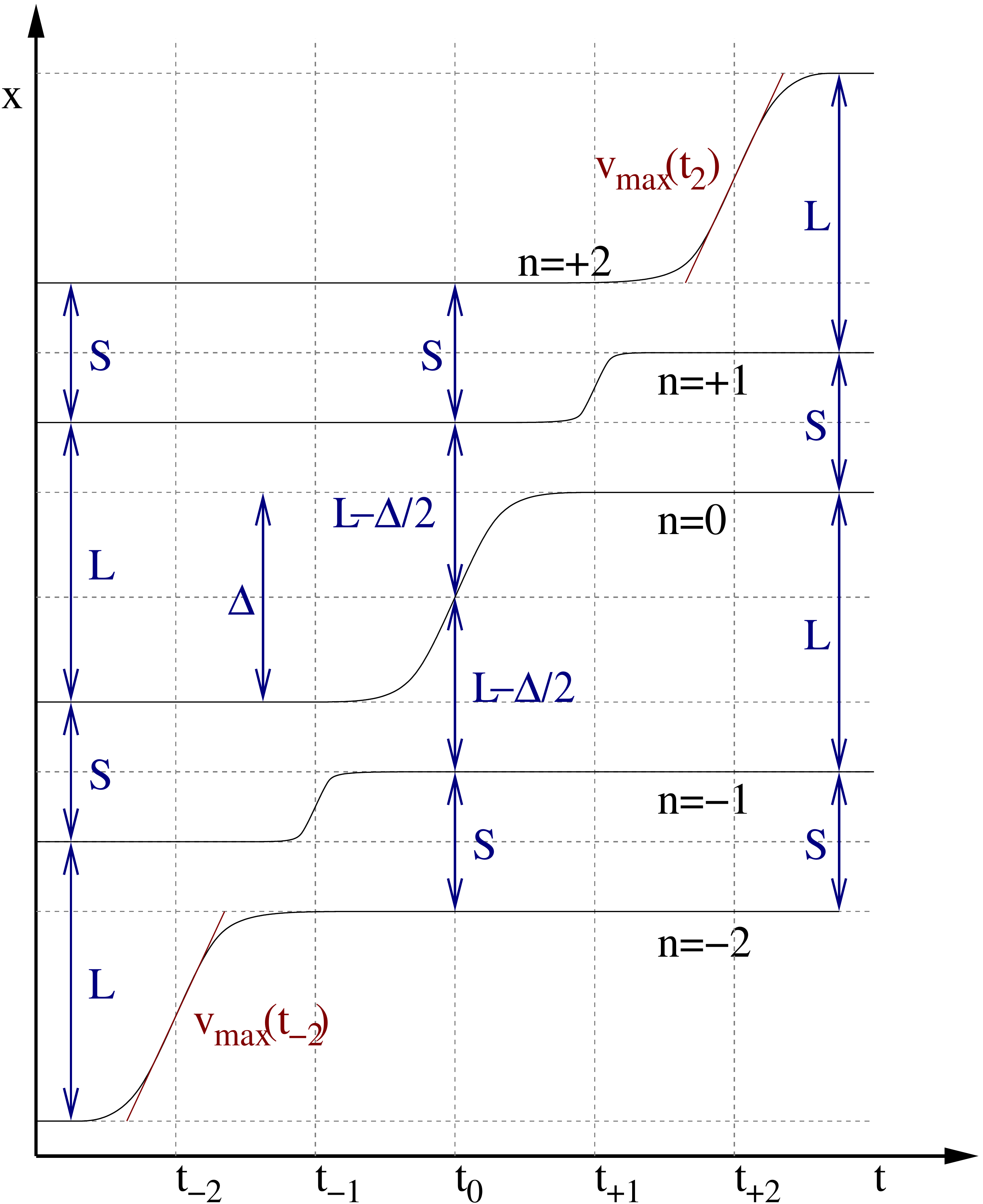}
		\caption{Schematic sketch of a propagating kink soliton: The particles jump by $\Delta$ or $\Delta+S-L$, the maximal
		particle velocities are shown as $v_{\mathrm{max}}(t_n)$.}
		\label{fig:skfliptraj}
	\end{figure}
\begin{figure}
	\subfigure[This typical kink soliton has been excited by assembling a periodic $LS$ chain, three particles, and the mirrored $LS$ chain
	and by giving the
	central particle the start velocity 15 ($\Delta=2$, $v_{\mathrm{max}}(0)=15$; see also figure \ref{fig:skfliptraj}).]{\label{fig:kinktraj}%
	\includegraphics[width=\linewidth]{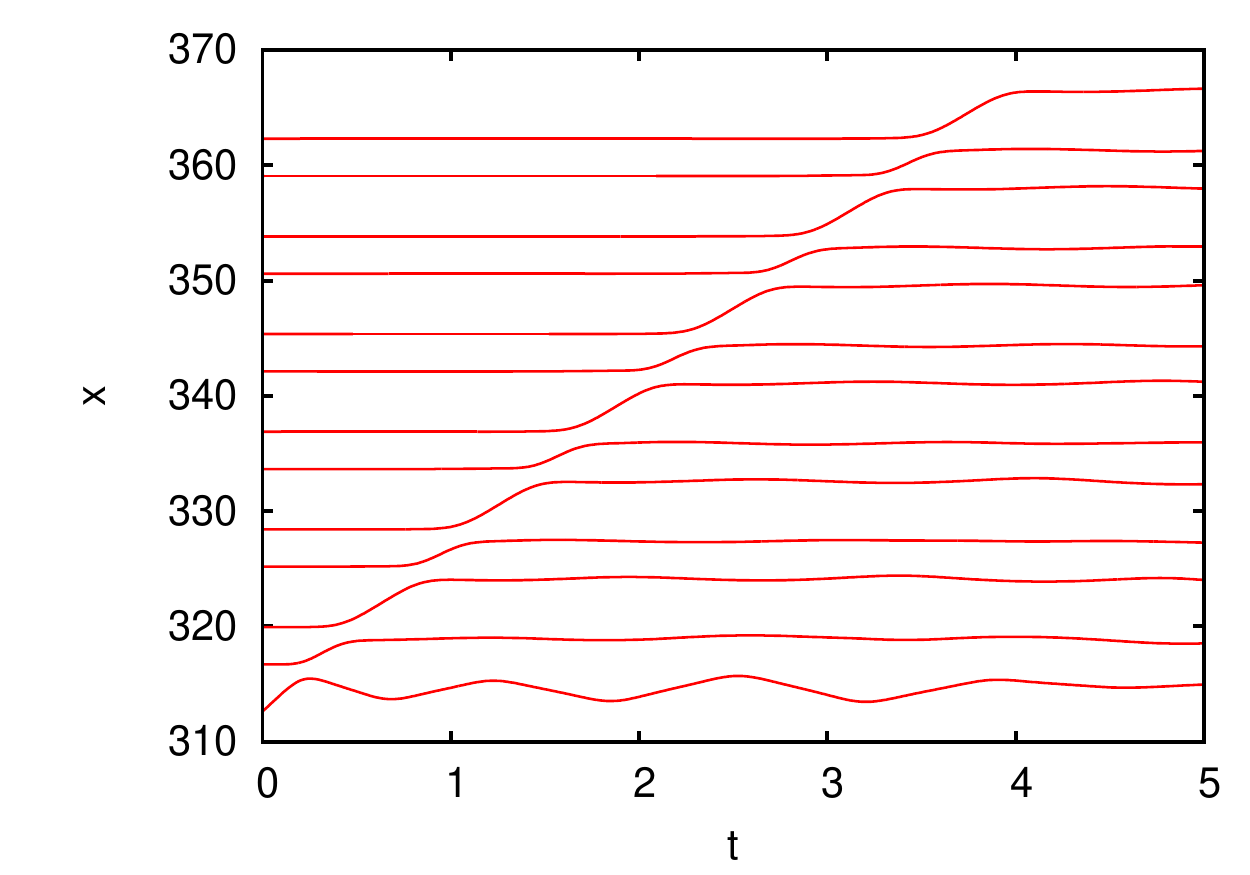}}%
	\\%
	\subfigure[The plot of the particle energy shows a kink soliton ($\Delta=4$, $v_{\mathrm{max}}(0)=7$)
	which loses energy by emitting phonon radiation. Consequently, the propagation velocity decreases until the kink decays.
	$t$ is the simulation time, $n$ the particle index in the chain.]{\label{fig:kinkrad}%
	\includegraphics[width=\linewidth]{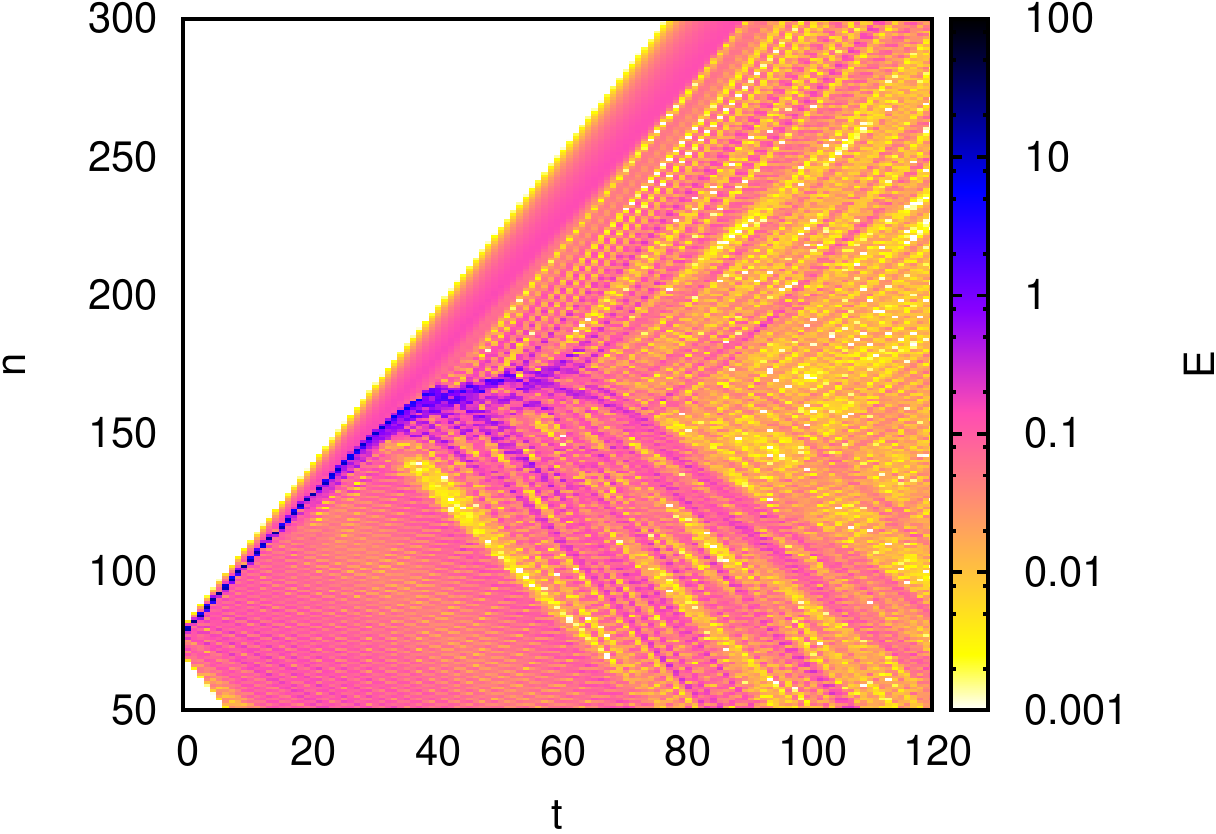}}%
	\caption{Propagating Kinks}
	\label{fig:kinkradtraj}
\end{figure}
\begin{figure}
	\centering
	\includegraphics[width=\linewidth]{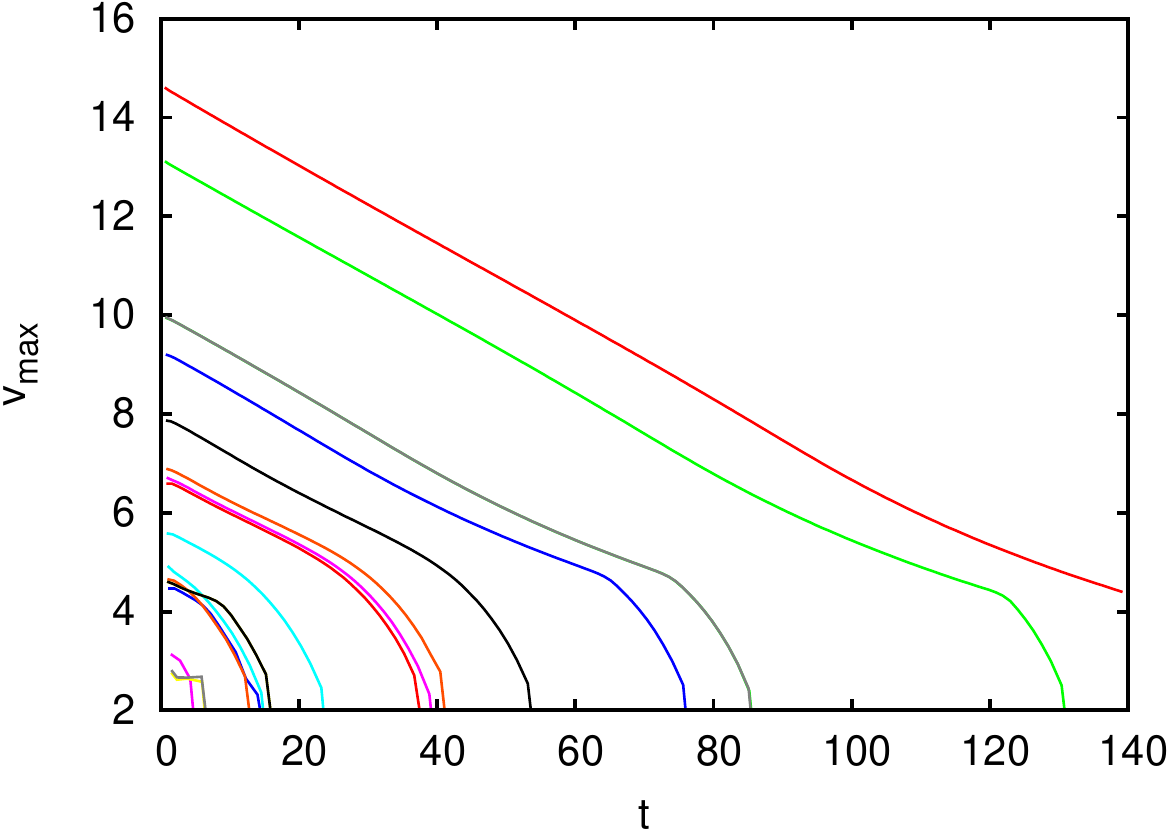}%
	\caption{Development of the maximal particle velocity $v_{\mathrm{max}}(t)$ of kink solitons.
	Independent of the excitation parameters ($\Delta$, $v_{\mathrm{max}}(0)$), the particle velocity decays nearly linearly in time
	until a critical velocity (about 5) is reached.}
	\label{fig:kinkv0}
\end{figure}

Figure \ref{fig:kinktraj} shows the typical trajectories in a chain with a propagating kink soliton.
For this simulation, initial conditions (figure \ref{fig:skfliptraj}) were chosen such that at $t=t_0$ the
particles of index $n=1,2,\dots$ form an $LS$ chain, as do the particles $n=-1,-2,\dots$, but particles
$n=1$ and $n=0$ are separated by $L-\frac{\Delta}2$, as are particles $n=-1$ and $n=0$. The tunable parameters
are the jump distance $\Delta$ and the initial velocity $v_{0}(t_0)=:v_{\mathrm{max}}(t_0)$ of particle 0.

The kink propagates by flipping atoms: For particles in $SL$ environment (seen in direction of kink propagation;
e.g. particle $n=0$ in the figure) one can observe jumps by $\Delta$, for particles in $LS$ environment
(particle 1) jumps by $\Delta+S-L$. During this process $LS$ environments become $SL$ environments and vice versa.
Of course, figure \ref{fig:skfliptraj} shows only an idealized, approximate picture of the time development of a
kink soliton. If $t=t_0$ is the moment of a snapshot of kink propagation with
maximal particle velocity $v_{\mathrm{max}}$, the jumping particle $n=0$ is very close to the center of the two neighbouring
particles $n=1,-1$. These particles also have finite (but low) velocities.
Because such deviations are not taken into account when initializing the system, the kinks typically need a short time to develop.

Because of the discrete lattice one expects that kinks radiate\cite{kinkrad}. In figure \ref{fig:kinkrad}, the total energies
$E_n(t)$ for particles $n$ at time $t$ are shown.
One can observe a kink emitting phonons. The flipping particle loses energy and
the kink
decelerates until it has decayed completely into phonons. During that process, the maximal particle
velocity decreases first approximately linearly in time, independent of the excitation
parameters $v_{\mathrm{max}}$, $\Delta$ at $t=0$,
and decays rapidly after reaching a critical velocity between 5 and 6, as figure \ref{fig:kinkv0} suggests.

The time dependence of the kink propagation velocity $v_\mathrm{kink}$ is qualitatively the same as $v_{\mathrm{max}}$
because the two quantities are almost linearly related 
independent of the excitation parameters.
So, in our system, there is no discontinuous variation of the kink's deceleration
as one expects for the Frenkel Kontorova chain, where two ``deceleration regimes'' exist\cite{fkvkink}.

Unfortunately, we could not find any analytical approximation describing the propagation of kinks.
It is evident that kinks can only exists at high energy scales compared to the energy barrier
of the potential (which is 1 in the partial interaction potential (\ref{eq:pot4})).

\subsection{Breathers}
\begin{figure}
	\subfigure[A rapidly decaying breather (type 1). The breather was constructed by combining a periodic $LS$ chain,
	five particles of distance $A$ with initial velocities $0,-3.5114,7.0228,-3.5114,0$, and another periodic $LS$ chain.]{\label{fig:brodd}%
	\includegraphics[width=\linewidth]{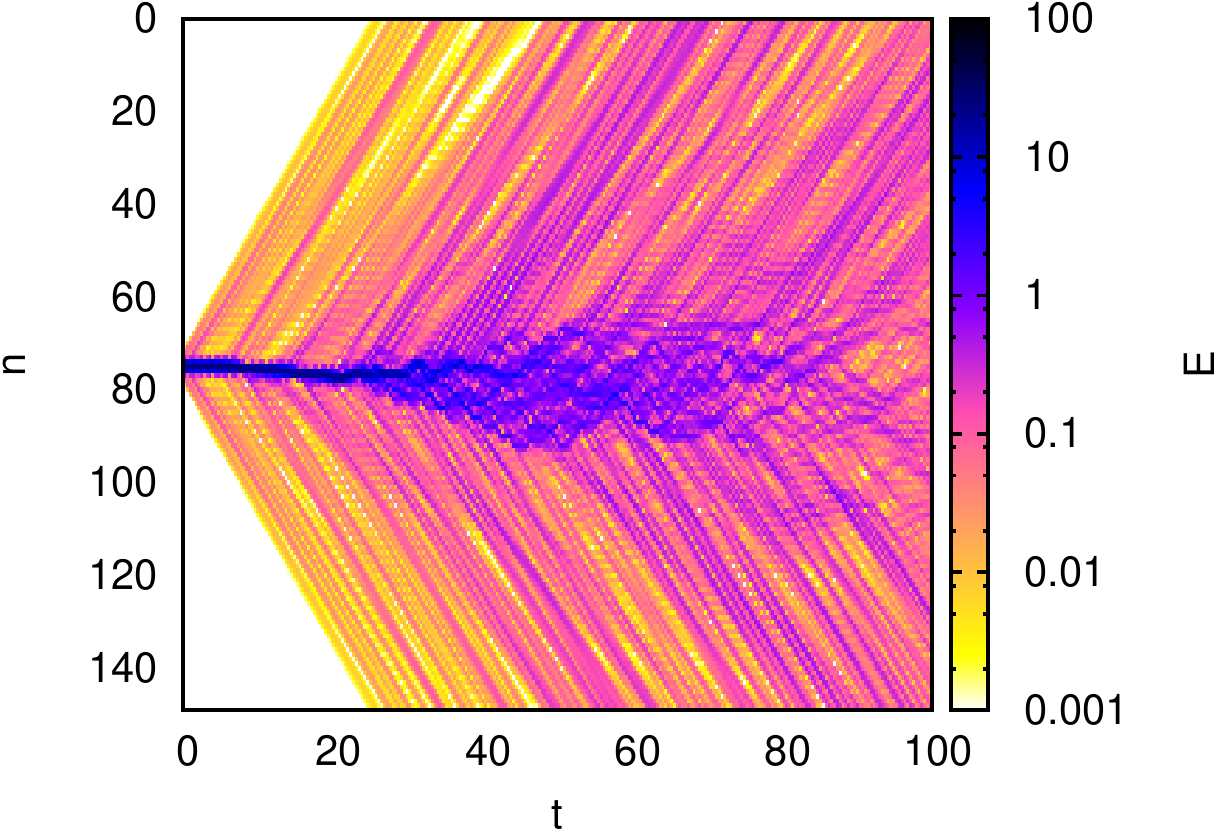}}%
	\\%
	\subfigure[Under ideal conditions (damped phonons, exact particle initialization)
	there exist long living breathers. This curve shows the energy profile of a breather of type 2 at $t=1.5\cdot10^6$.]{\label{fig:brprof}%
	\includegraphics[width=\linewidth]{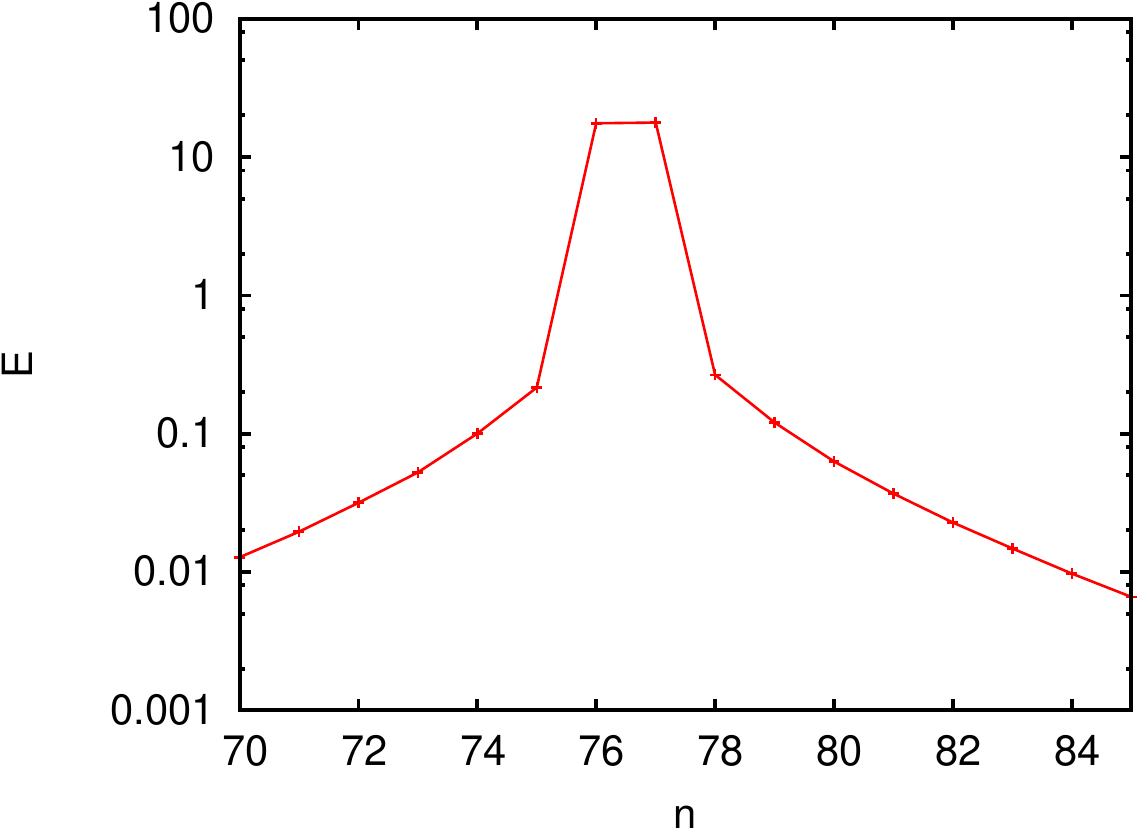}}%
	\caption{Simulated Breathers: Energy}
\end{figure}
\begin{figure}
	\centering
	\includegraphics[width=\linewidth]{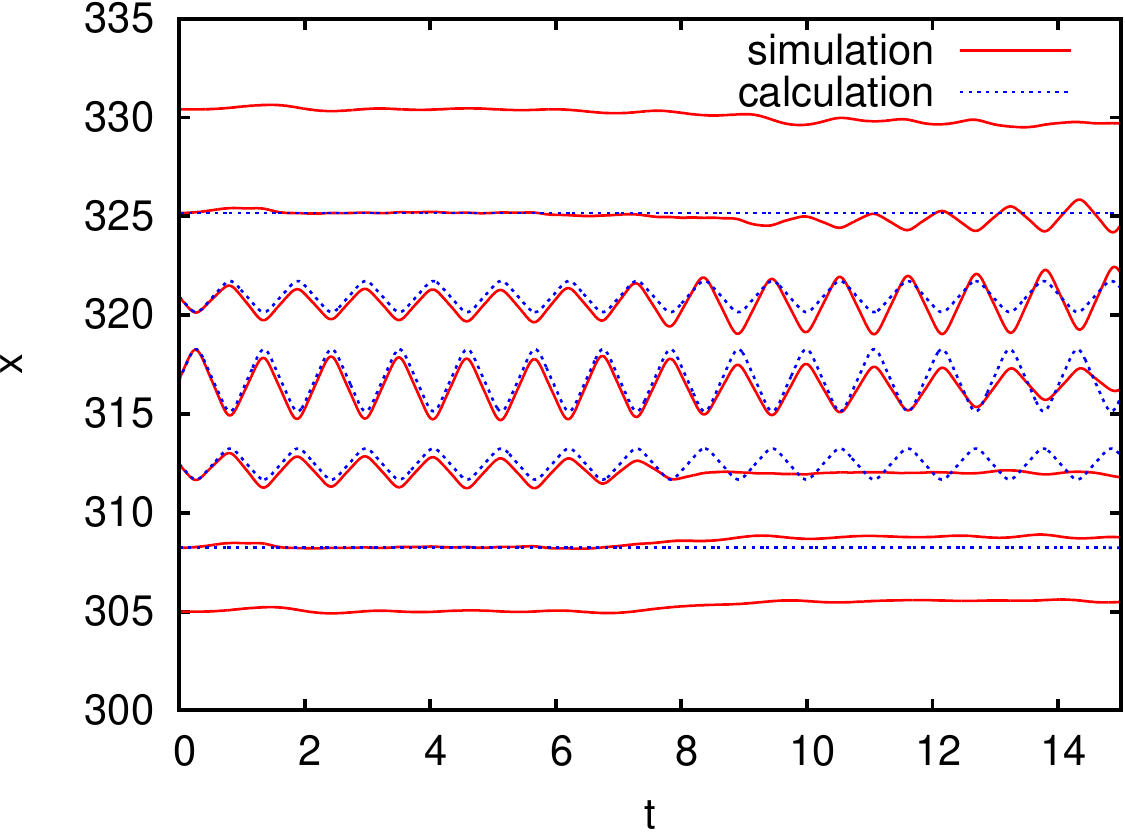}
	\caption{Comparing the simulated trajectories of a type 1 breather and the analytical approximation.
	The motion is described well for short times.}
	\label{fig:brana}
\end{figure}

Breathers are localized oscillatory modes. Numerical experiments (figure \ref{fig:brodd}) show that they are unstable.
They can be destroyed by interaction with phonons or by tiny changes in the particle configuration

We observed two types of breathers: One (type 2) basically consists of two neighbouring particles oscillating in opposite directions.
The other (type 1) consists of a central oscillating particle and its oscillating neighbours. For the latter (type 1), our simulations suggest
the following analytical approximation:
\begin{subequations}
\begin{align}
	x_{-2}(t) &= c_1         \\
	x_{-1}(t) &= c_2 - X(t)  \\
	x_0(t)    &= c_3 + 2X(t) \\
	x_{+1}(t) &= c_4 - X(t)  \\
	x_{+2}(t) &= c_5
\end{align}
\end{subequations}

Conservation of energy and
$V(x) = ax^4+bx^2$, $c_3 = 0$, $c_4 = A$, $c_5 = 2A$, $c_2 = -A$, and $c_1 = -2A$ (this is due to numerical observations)
lead to
\begin{equation}
	E = 3M \dot{X}(t)^2 + 164aX(t)^4+20bX(t)^2
\end{equation}
The solution for the standard double well potential ($a=1$ and $b=-2$) is
\begin{align}
	X(t) &= \pm\sqrt{\frac{E}{3M}} \Omega^{-1} \Sd\left(\Omega(t-t_0)|m\right)\label{eq:brx1}
	\\
\intertext{with}
	\Omega &= \sqrt{\frac{40}{3M}\sqrt{1+\frac{41}{100}E}}
	\\
	m &= \frac12 \left(1+\frac1{\sqrt{1+\frac{41}{100}E}}\right)
\end{align}

The other type of breather (type 2) can be described by
\begin{subequations}
\begin{align}
	x_{-\frac32}(t) &= -\frac32 A \\
	x_{-\frac12}(t) &= -\frac12 A - X(t) \\
	x_{+\frac12}(t) &= +\frac12 A + X(t) \\
	x_{+\frac32}(t) &= +\frac32 A
\end{align}
\end{subequations}
which has this solution for our potential:
\begin{align}
	X(t) &= \pm\sqrt{\frac{E}{M}} \Omega^{-1} \Sd\left(\Omega(t-t_0)|m\right)
	\\
\intertext{with}
	\Omega &= \sqrt{\frac{12}{M}\sqrt{1+\frac12 E}}
	\\
	m &= \frac12 \left(1+\frac1{\sqrt{1+\frac12 E}}\right)
\end{align}

Although this localized ansatz is no stable solution,
numerical simulations show good agreement for short times (figure \ref{fig:brana}). If the system is set up
carefully and phonons, which would destroy the breather, are damped out, the system stabilizes
when the neighbouring atoms oscillate with very low amplitudes. Figure \ref{fig:brprof}
e.g. shows the energy profile in the chain at $t=1.5\cdot10^6$. The high value of $t$ demonstrates how stable
breathers can be under special conditions.

\section{Dynamic Fibonacci chain}\label{sec:dfc}

In this section we present simulations of the dynamic Fibonacci chain, i.e.
a chain of atoms whose initial positions are that of particles in the Fibonacci chain using
our usual interaction potential with $a=1$ and $b=-2$.

We set the initial velocities to random numbers that are normally distributed and allow us to
simulate finite temperatures.

We prefer a temperature regulation by setting $\langle{}E\rangle$ to a defined value and waiting
for the system to reach an equilibrium state,
since numerical thermostats
can change the behaviour of the system
and because there is no simple relation between $T$ and $\langle{}E\rangle$.

\subsection{Ground state}

If the length of the chain is a sum of integer multiples of $L$ and $S$, then all particles can
sit in potential minima, which is a ground state of the system. For a fixed chain length or fixed
periodic boundary conditions, the total number of $L$ and $S$ remains fixed because of their irrational ratio.
All rearrangements by flips ($LS$ to $SL$ and vice versa) lead to energetically equivalent states.

The Fibonacci chain is a special case defined by $p(L)=\tau p(S)$, where $p(L)$ and $p(S)$ are the frequencies of occurrence of $L$ and $S$
segments in the chain. Furthermore, the Fibonacci chain is the most uniform distribution of these two tile types within the class of all chains
with the ratio $\tau$. At finite temperatures, the chain will become disordered by flips, and defects like $SS$ will form
that are not present in the original Fibonacci chain. In thermodynamical equilibrium, the Fibonacci chain will approach a random tiling.

For low energies, the particles basically feel the parabolic potential that results from the Taylor
expansion of the potential around its minima (i.e. $\langle{}E_\mathrm{kin}\rangle\approx \frac12\langle{}E\rangle$).
The quartic term only dominates in the high energy regime $E>1$ ($\langle{}E_\mathrm{kin}\rangle\approx \frac23\langle{}E\rangle$).
As we will see, the nonparabolicity leads to significant deviations of the specific heat from Dulong-petit law.

In the following two sections, we will study the randomization of the Fibonacci chain and its heat capacity.

\subsection{Randomization}
	\begin{figure}
		\subfigure[Average frequency of traversals of potential centers dependent on simulation time
		for different average energies.
		This frequency is a measure for the number of $LS$ environments in the chain
		which changes due to the transition of the Fibonacci chain to a random tiling.
		The fit of the curve for $\langle{}E\rangle=0.15$ to the exponential law given by equation (\ref{eq:explaw}) is quite good.]{\label{fig:avgs10}%
		\includegraphics[width=\linewidth]{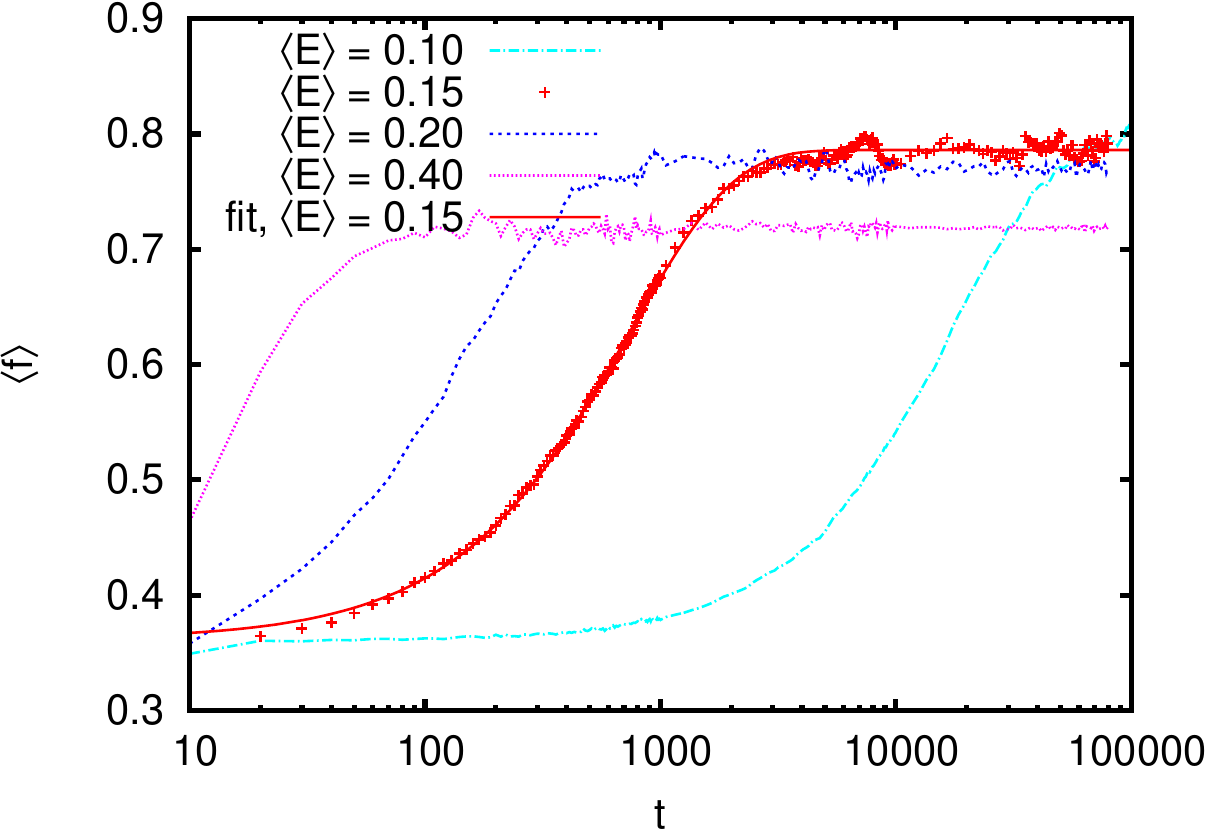}}%
		\\%
		\subfigure[Arrhenius plot of the inverse randomization time, using $\langle{}E\rangle$ as a measure of temperature]{\label{fig:avgs10fit}%
		\includegraphics[width=\linewidth]{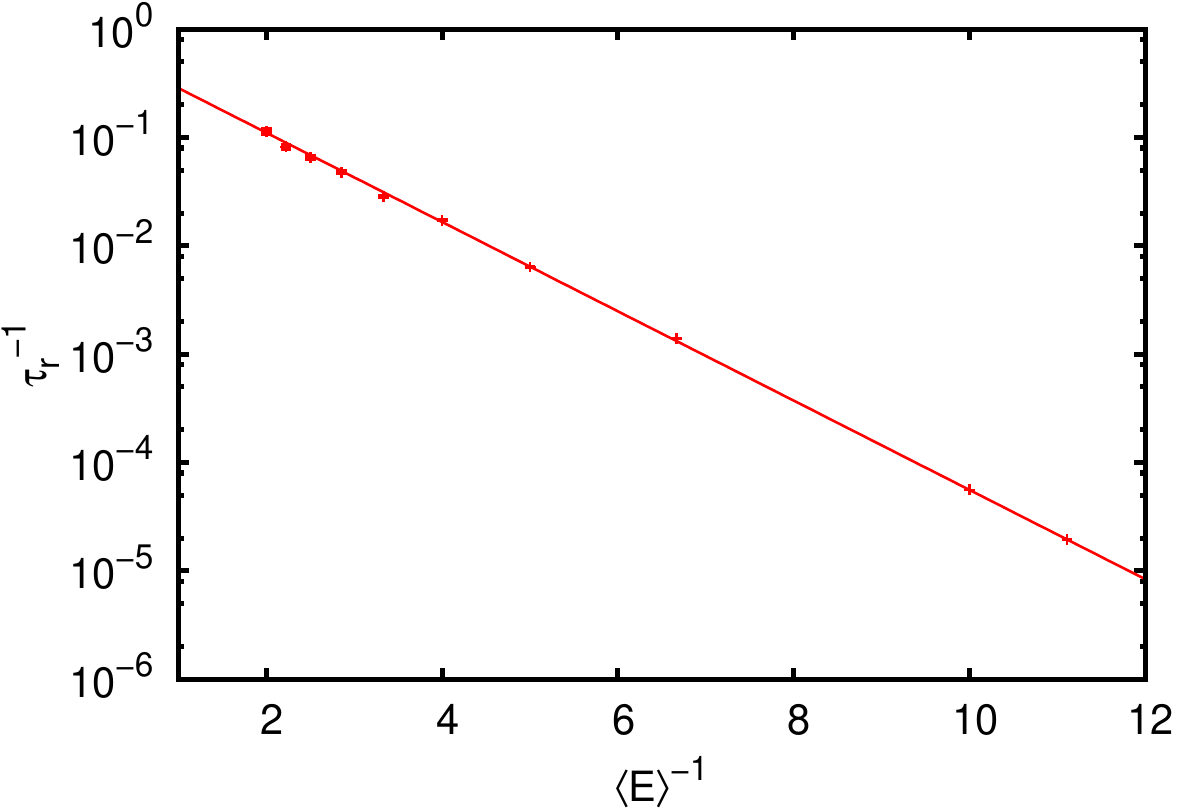}}%
		\caption{Relaxation of the dynamic Fibonacci chain}
	\end{figure}

While most observables we studied in the system showed quite plausible behaviour throughout our simulations (e.g. the flip frequency discussed below),
we observed strange transitions
of behaviour for such observables that depend on the number of $LL$, $LS$ and $SS$ environments in the system.
The reason are structural changes in the chain that can be observed for high temperatures or long simulation times, when
the Fibonacci chain has evolved to a random tiling.

The Fibonacci chain has another ratio of the possible environments
\begin{equation}
p(SS):p(LS):p(LL) = 0:\tau^{-1}:\tau^{-2}
\end{equation}
with
\begin{equation}
p(L):p(S) = \tau^{-1}:\tau^{-2}
\end{equation}
than the random tiling
\begin{equation}
p(SS):p(LS):p(LL) = \tau^{-4}:2\tau^{-3}:\tau^{-2}
\end{equation}
Therefore observables such as the average frequency of particle transitions through the potential centers
(zero for $LS$ environments because the particles are trapped in one half of
the double wells) change to an equilibrium value until the random tiling is reached,
following an exponential law
\begin{equation}
	\label{eq:explaw}
	f(t) = f_{\infty} + (f_{0}-f_{\infty}) \Exp{-t/\tau_{\mathrm{r}}}
\end{equation}
with the ``randomization time'' $\tau_{\mathrm{r}}$, see figure \ref{fig:avgs10}.

Figure \ref{fig:avgs10fit} shows, that this randomization time follows an Arrhenius law
\begin{equation}
\label{eq:Ereconst}
\tau_{\mathrm{r}}(T)^{-1} \propto \Exp{-E_{\mathrm{r}}/kT}
\end{equation}
where we find the ``relaxation energy'' $E_{\mathrm{r}}\approx 0.95$.

\subsection{Specific heat}
	\begin{figure}
		\centering
		\includegraphics[width=\linewidth]{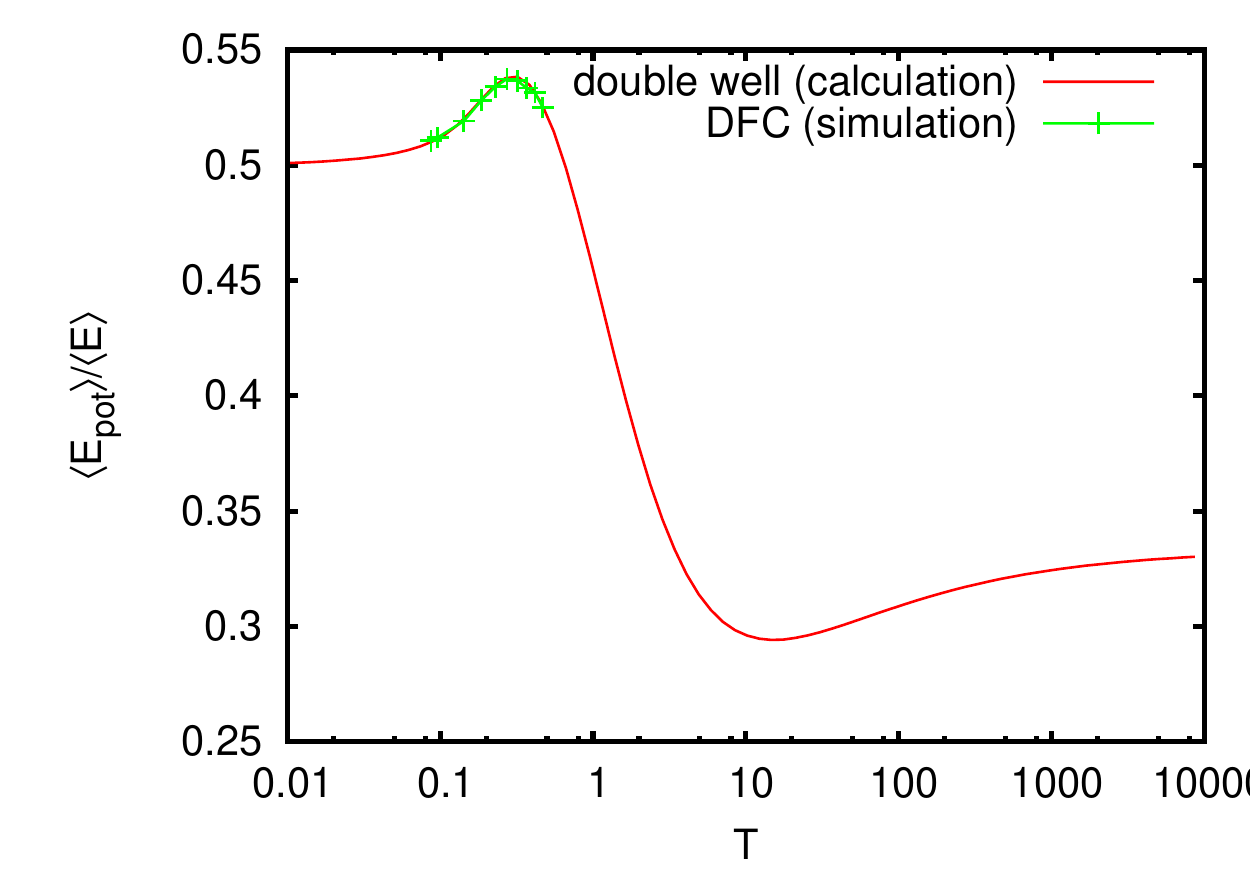}
		\caption{The ratio of average potential energy and average total energy in the chain is the same
		as for non interacting particles in a double well potential. Thermodynamic properties are
		governed by the anharmonic potential.}
		\label{fig:dfcpoten}
	\end{figure}
	\begin{figure}
		\centering
		\includegraphics[width=\linewidth]{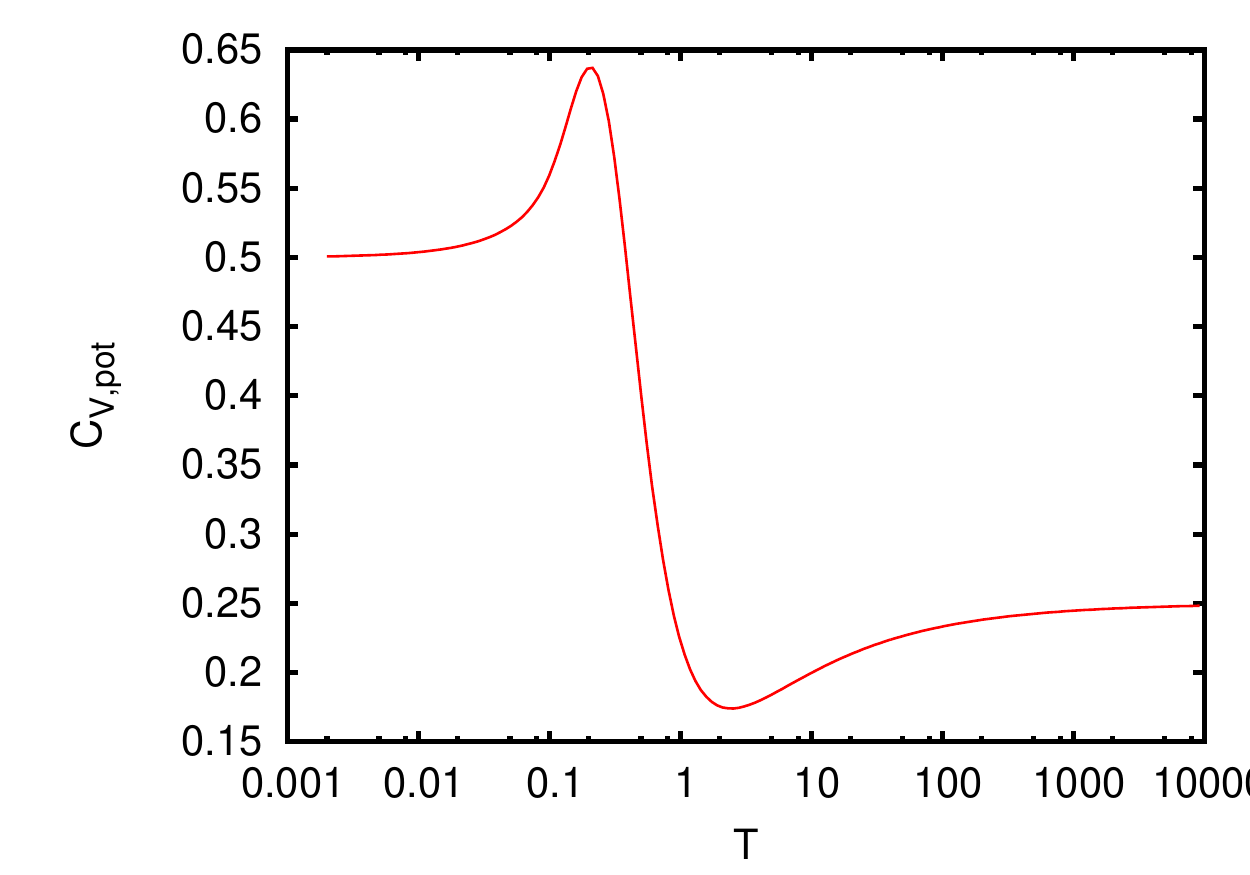}
		\caption{Temperature dependence of the specific heat of non interacting particles in a double well potential
		obtained by analytical calculation.}
		\label{fig:anhheatcap}
	\end{figure}
	Because of the degenerate ground states, flips do not contribute to entropy and hence, to the specific heat,
	which is solely determined by the interaction potential. We will prove this statement in the following.

	Figure \ref{fig:dfcpoten} shows for the ratio $\langle E_\mathrm{pot}\rangle/\langle E\rangle$ that thermodynamical
	properties of the dynamic Fibonacci chain agree remarkably with those of particles in a double well potential. It also demonstrates
	the transition from harmonic behaviour ($\langle{}E_\mathrm{pot}\rangle\approx \frac12\langle{}E\rangle$) to quartic behaviour ($\langle{}E_\mathrm{pot}\rangle\approx \frac13\langle{}E\rangle$) very well. In the relevant temperature range $0.1 \lessapprox T \lessapprox 1$ for our simulations
	of the Fibonacci chain we are close to parabolic behaviour but already expect effects of anharmonicity.

	Therefore, we use the analytically accessible system of noninteracting particles in a double well potential to
	determine 
	the specific heat: The contribution of the potential energy can be calculated from
	\begin{align}
	C_\mathrm{V,pot}&=\frac{\partial \langle E_\mathrm{pot}\rangle}{\partial T}\\
	&=\frac{1}{kT^2}\left(\langle E_\mathrm{pot}^2\rangle-\langle E_\mathrm{pot}\rangle^2\right)
	\end{align}
	using
	\begin{align}
	\left\langle E_\mathrm{pot}\right\rangle&=E_0 \frac{A^-_2(\vartheta)-2A^-_1(\vartheta)}{A^-_0(\vartheta)}\\
	\left\langle E_\mathrm{pot}^2\right\rangle&=E_0^2 \frac{A^-_4(\vartheta)-4A^-_3(\vartheta)+4A^-_2(\vartheta)}{A^-_0(\vartheta)}
	\end{align}
	with $\vartheta:=kT/E_0$ and the integrals
	\begin{align}
	A^\pm_n(a)
	&:=\int\limits_{-\infty}^{\infty}x^{2n} \Exp{-\frac{x^4\pm 2x^2}{a}} \,\dx
	\\
	&=\left(\frac{a}{2}\right)^{\frac{2n+1}{4}} \GAMMA\left(n+\frac12\right)
		\Exp{\frac{1}{2a}}U\left(n,\pm\sqrt{\frac{2}{a}}\right)
	\end{align}
	where $U$ is a parabolic cylinder function\cite{abrsteg}.
	The plot of this function (figure \ref{fig:anhheatcap}) shows a rather interesting transition between
	harmonic and quartic regime: With increasing temperature $C_\mathrm{V,pot}$ rises from the parabolic $\frac12 k$ before
	approaching the quartic limit $\frac14 k$, which has already been discussed in literature.\cite{dwellheat}
	As the contribution of the kinetic energy $C_\mathrm{V,kin}=\frac12k$ is constant, the total specific heat
	rises from the parabolic value $k$ before declining and finally approaching quartic $\frac{3}{4}k$ from beneath.

	Edagawa et al.\cite{edagawaheat,prekulheat} observed that for high temperatures, $C_\mathrm{V}$ of $\mathrm{Al}_{63}\mathrm{Cu}_{25}\mathrm{Fe}_{12}$ rises
	above $3k$ which one expects from Dulong Petit.
	It is interesting that even for our simple model an increase of the specific heat by 14\% can be observed,
	which is only caused by the interaction potential.
	This observation is in good agreement with the results of Grabowki et al.\cite{grabheat} which suggest that anharmonic
	interaction potentials cause high values of specific heat even for elementary fcc metals as Aluminium.

	As seen above, no contribution of phasons to the specific heat is observed at
	high temperatures. W\"{a}lti et al. have measured an excess specific heat at
	low temperatures and assume its origin to lie in nonpropagating lattice
	excitations\cite{lattexc}. In our system we expect no influence of phasons on the low
	temperature specific heat, too, due to the low energy cutoff for the phason
	flips (Fig. 15). Low energy nonacoustic phonons are not present in the DFC, but
	in the asymmetric Fibonacci chain (AFC, Engel et al.\cite{dfcpaper}), which we
	are not dealing with here. The AFC is characterized by a bias in the double
	well which causes anticrossings in the many branches of the dynamical structure
	factor and leads to low lying flat bands (Fig. 15 of Engel et al.\cite{dfcpaper}).
\subsection{Flip energy}
	\begin{figure}
		\subfigure[For an ``autonomous flip'' the flipping particle has enough kinetic energy to cross the energy barrier
		           in the local potential landscape.]{\label{fig:flipaut}%
		           \includegraphics[width=1.0\linewidth]{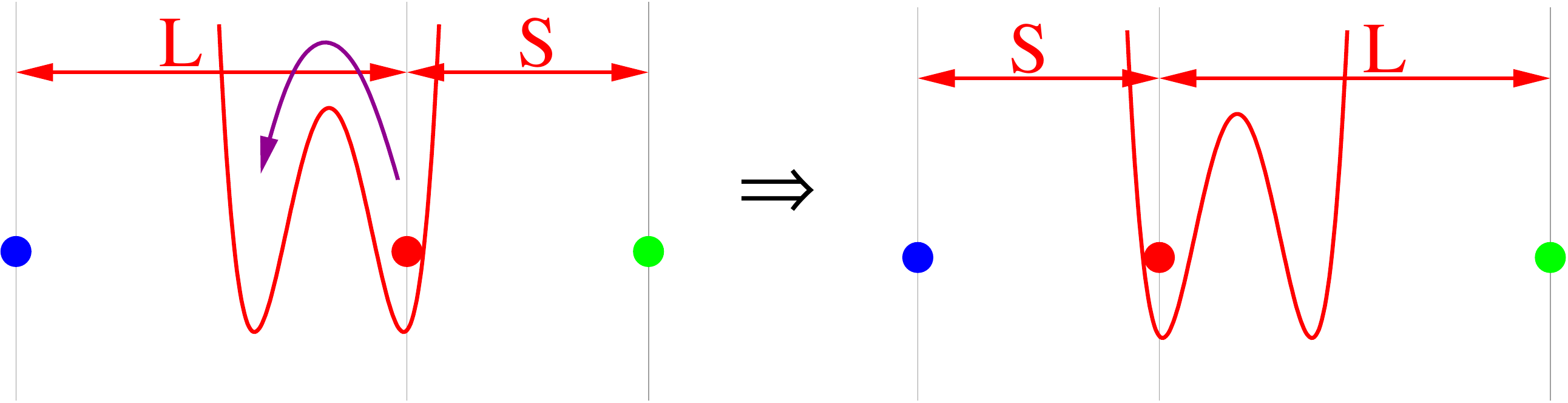}}%
		\\%
		\subfigure[For an ``induced flip'', the ``flipping'' particle does not have to move at all: The motion of
		           the neighbouring particles can change the potential landscape in such a way that the energy barrier disappears at one side of the particle
		           and reappears on the other side.]{\label{fig:flipind}%
		           \includegraphics[width=1.0\linewidth]{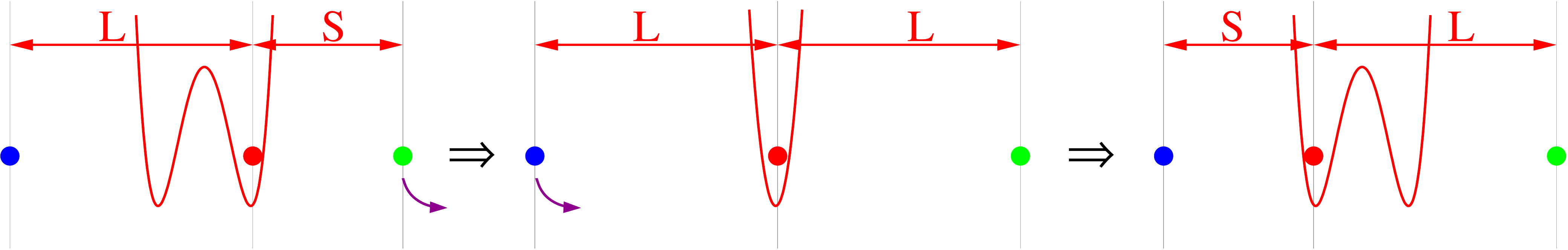}}
		\caption{Flip mechanisms}
		\label{fig:flipmech}
	\end{figure}
	\begin{figure}
		\subfigure[The particle energies of atoms in the moment of a flip show two interesting features:
		A lower threshold energy $E \approx 0.9$ and a peak at $E\approx 2$.]{\label{fig:flipen-fp-tot}%
		\includegraphics[width=\linewidth]{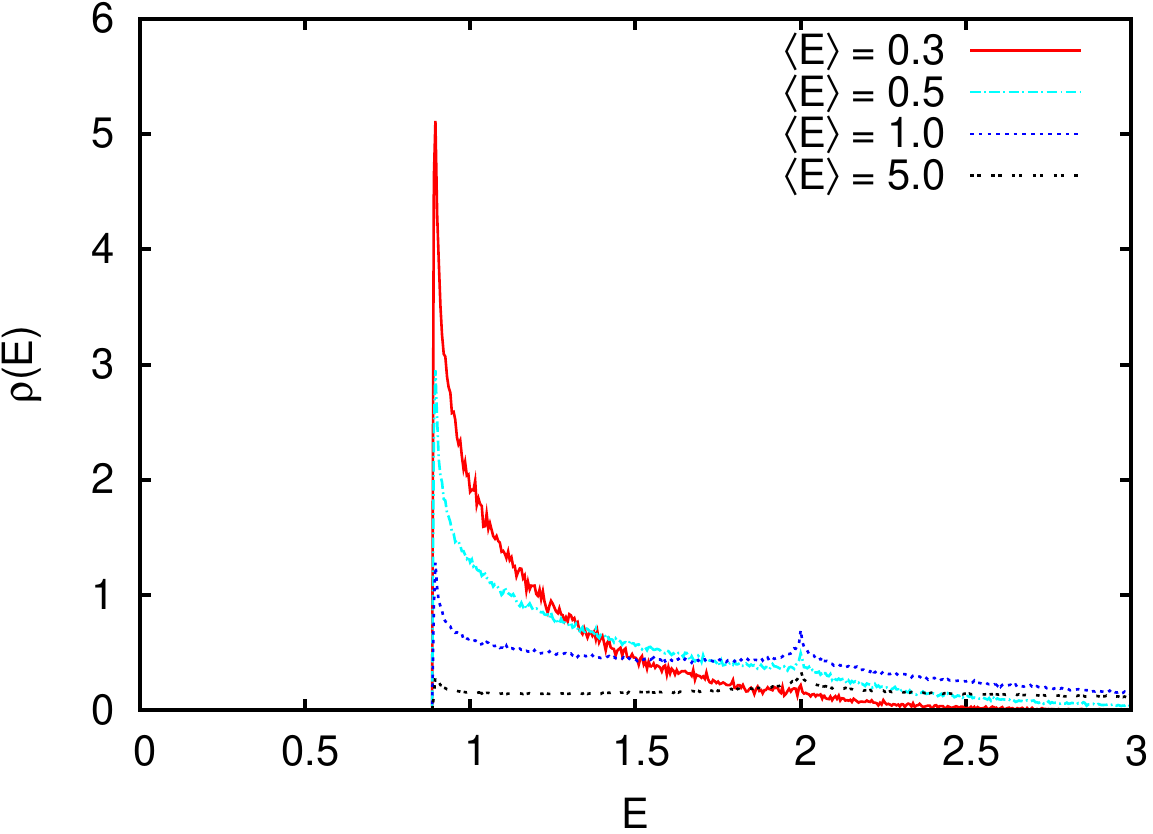}}%
		\\%
		\subfigure[The energy distribution for autonomous flips shows the threshold energy and the peak even better because
		time and energy of induced flips are somewhat difficult to define.]{\label{fig:flipen-fp-notass}%
		\includegraphics[width=\linewidth]{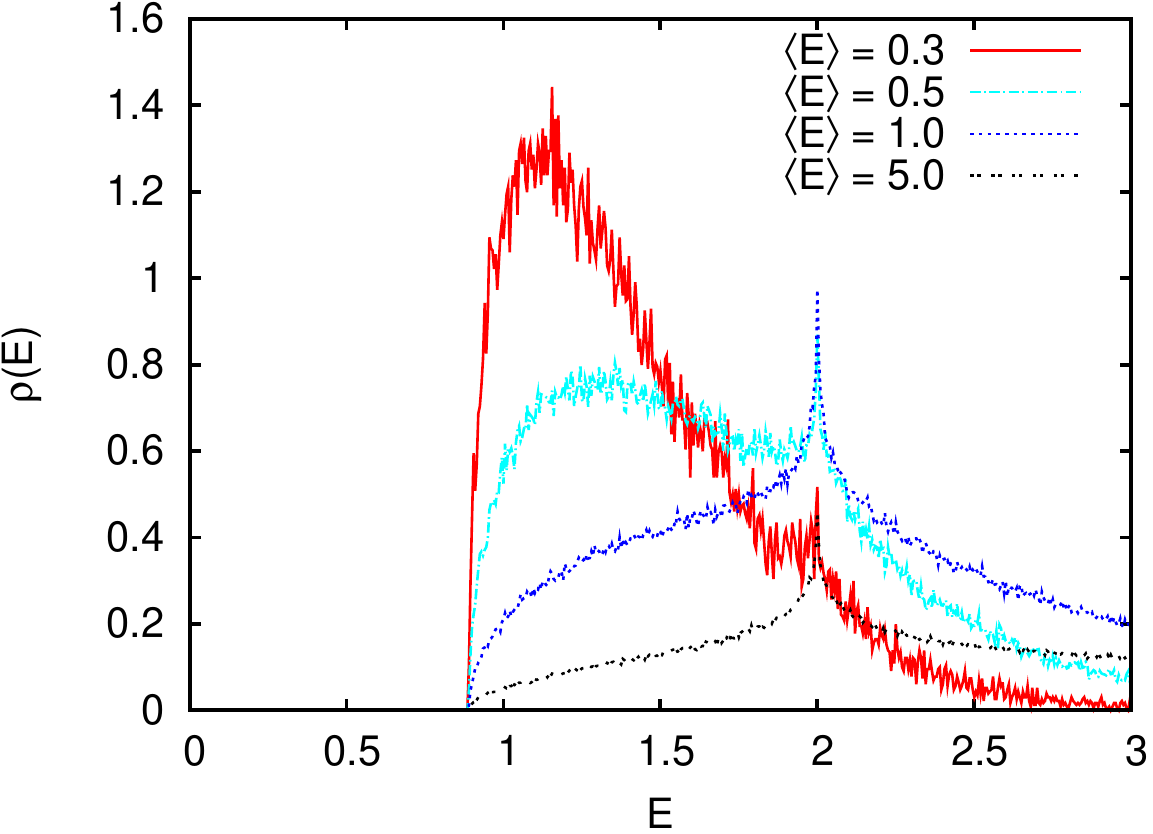}}%
		\caption{Energy distribution of flipping atoms}
	\end{figure}
To understand flips better, we calculated the energy distribution of particles
in the moment a flip occurred.

There are two types of flip processes: ``Autonomous flips''
when atoms have enough energy to cross the
potential barrier (figure \ref{fig:flipaut}). For ``induced flips'' on the other hand, the flipping particle does not
have to move at all: Particle motions of the next neighbours change the local energy landscape and cause
the energy barrier to disappear. When a new barrier rises at the other side of the particle, the
particle has completed a flip
(figure \ref{fig:flipind}).

We calculated the average energies of
particles in the moment of
autonomous flips
or of
induced flips (here
the ``moment of the flip'' is somewhat arbitrary, one can choose any point in the period of time when no barrier exists;
we chose the moment when the energy barrier reappears).
The particle energies were calculated as sum of kinetic energy and the local potential energy equation (\ref{eq:neighbourpot})
shifted by $2\cdot\frac{b^2}{4a}$ which causes the minimal energy of the double wells
to be zero.

The probability distributions (figures \ref{fig:flipen-fp-tot} and \ref{fig:flipen-fp-notass}) display two prominent features: There
is a sharp peak at $E=2$, which is exactly the energy of the potential barrier.
It is explained by the vanishing slope of $V(x)$ there and the corresponding high density of states.

The other interesting point is the sharp cutoff at $E \approx 0.9$ which exists for both autonomous and induced flips.
This lower threshold energy can be explained as the minimal potential energy of the energy barrier, i.e. the potential energy
at the potential center in the case where the neighbouring atoms just reach the critical distance (equation (\ref{eq:dcrit})).
This ``minimal flip energy'' is calculated to $E_\mathrm{crit}=\frac89$.

It is remarkable that this minimal flip energy is close to the relaxation energy of
equation (\ref{eq:Ereconst}).

\subsection{Energy transport and flips}

\begin{figure}
	\centering
	\includegraphics[width=\linewidth]{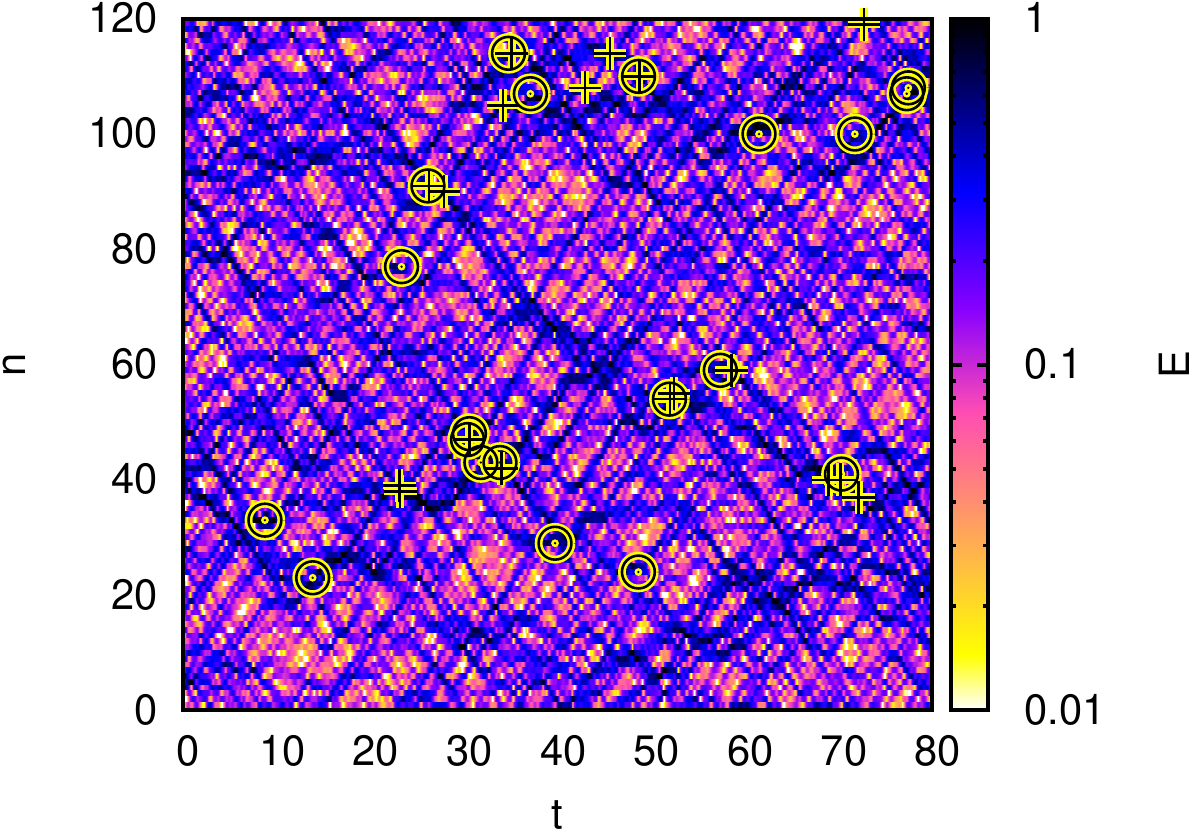}
	\caption{Particle energies in the dynamic Fibonacci chain at $\langle E\rangle=0.2$. Energy is transported
	as uniform motion of high energy regions which appear as straight lines in this graph. Because the energy in
	these regions can exceed the threshold energy for flips, induced flips (circles: $\odot$) and autonomous flips
	(crosses: $+$) can be found along these lines.}
	\label{fig:oren02}
\end{figure}
\begin{figure}
	\includegraphics[width=\linewidth]{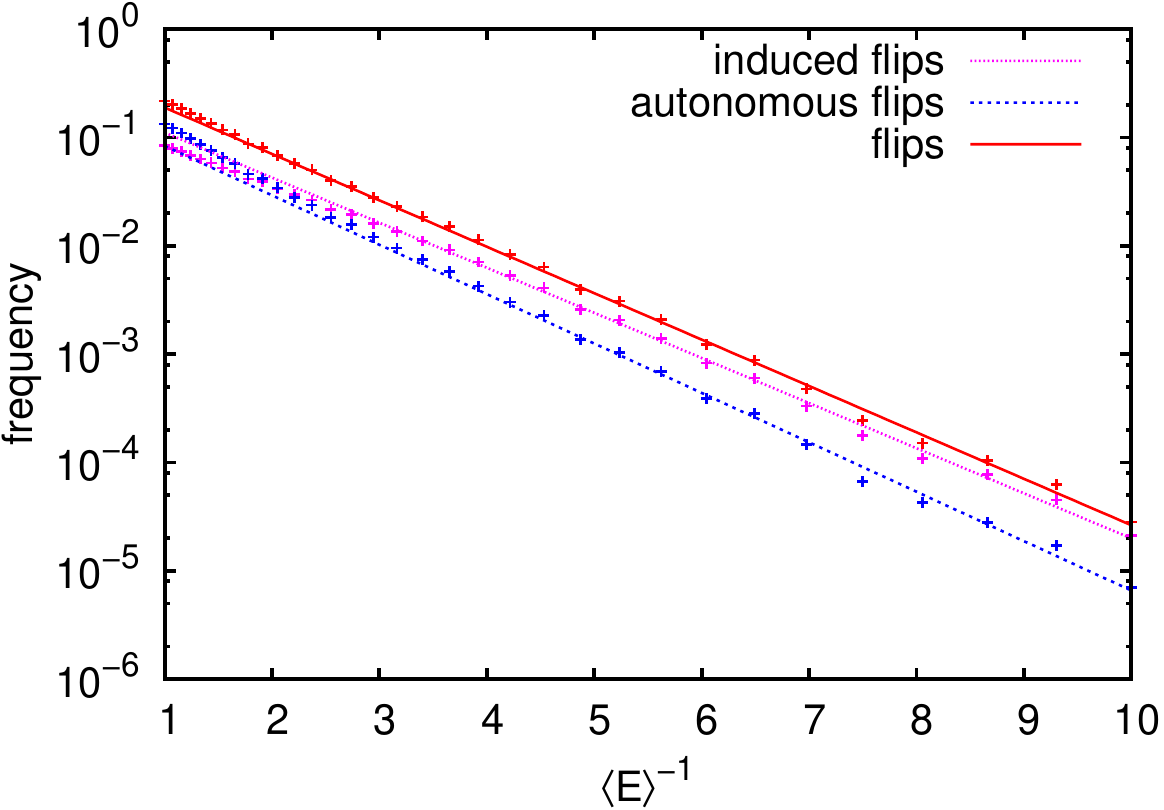}%
	\caption{For low temperatures, the Arrhenius plot of the average flip frequency in the DFC shows
	the expected behaviour for all flip types.}
	\label{fig:flipfreqlowfit}
\end{figure}
When studying energy transport in the chain at thermodynamical equilibrium and at low energies (i.e. energies lower
than the potential barriers), we could not observe any prominent soliton modes
in accordance with the results of section \ref{sec:ls} on the $LS$ chain.

On the other hand we could find low energy modes which can be described as regions of high energies propagating with
constant velocity (they appear as straight lines in figure \ref{fig:oren02}). Because there is a lower treshold energy
for flips, one expects that flips occur essentially along these propagating energy packets.

To verify this assumption, we marked the flips in the figure: Autonomous flips are represented by crosses ($+$), induced flips
by circles ($\odot$). As it turns out, flips are indeed concentrated close to these modes.

When we increase temperature (i.e. $\langle E\rangle$), more and more of these high energy regions
cross the energy threshold and cause a rapidly growing number of flips. Consequently, the flip frequency (number of flips per
unit time and particle) is expected to follow an Arrhenius law.

Figure \ref{fig:flipfreqlowfit} shows the Arrhenius plot of the rapidly inreasing frequencies we measured in our simulated system. As the straight lines fit our data nicely, the plot confirms our assumption. The slopes suggest
activation energies 0.96, 1.05, 0.98 for induced, autonomous, and all flips which are fairly close to the minimal flip
energy.

\subsection{Solitons in the dynamic Fibonacci Chain}

Next, we investigate the response of the chain to local excitations. All
particles start on their equilibrium positions with zero velocity. Then, at
time $t_{0}$, one particle $n_{0}$ is `kicked' by setting
$v_{n}(t_{0})=v_{0}\delta_{n,n_{0}}$.

\begin{figure}
  \subfigure[Interaction of two solitons]{\label{fig:dfcmodeskink}%
  \includegraphics[angle=-90,width=\linewidth]{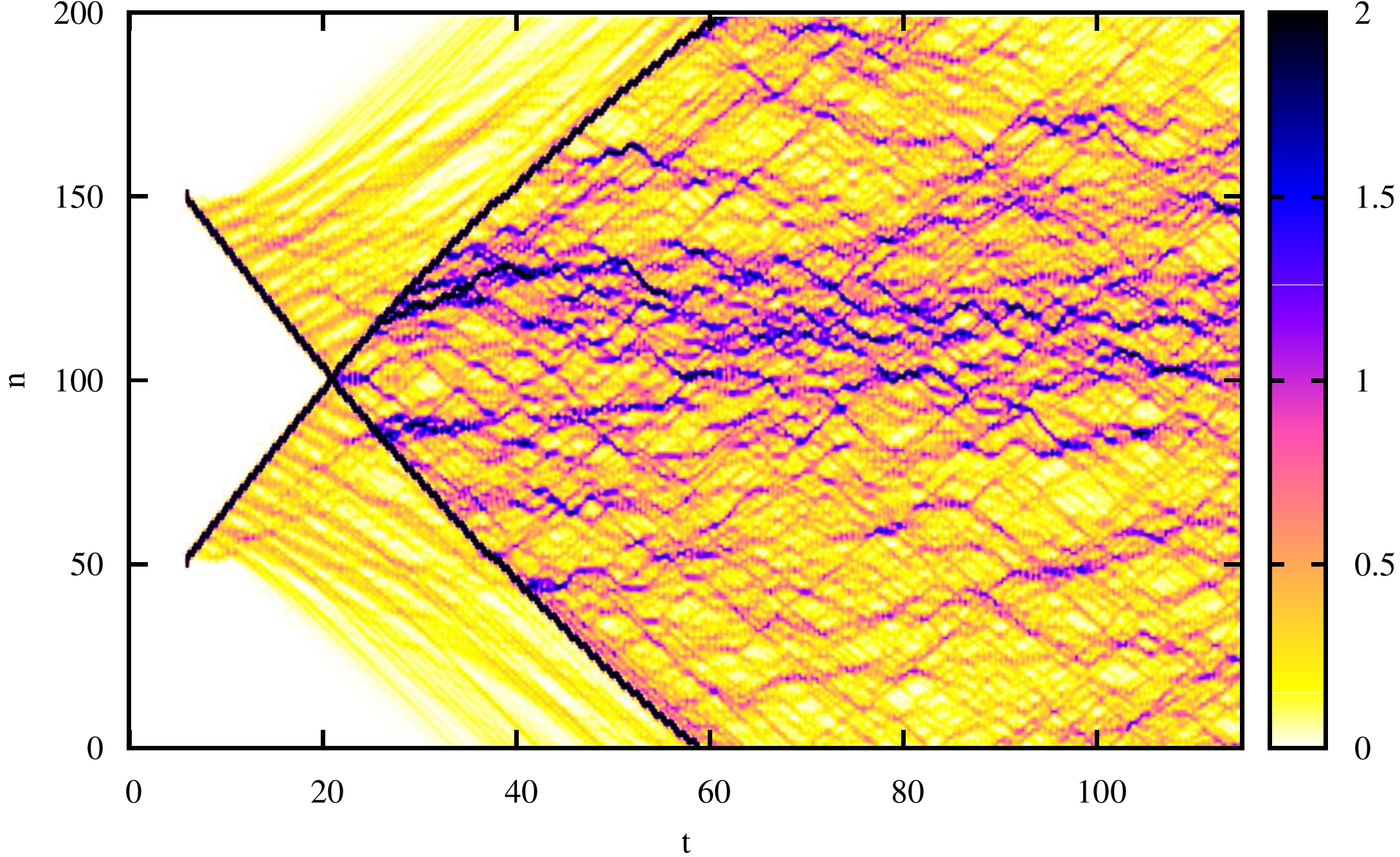}}%
  \\%
  \subfigure[A stationary breather]{\label{fig:dfcmodesbreather}%
  \includegraphics[angle=-90,width=\linewidth]{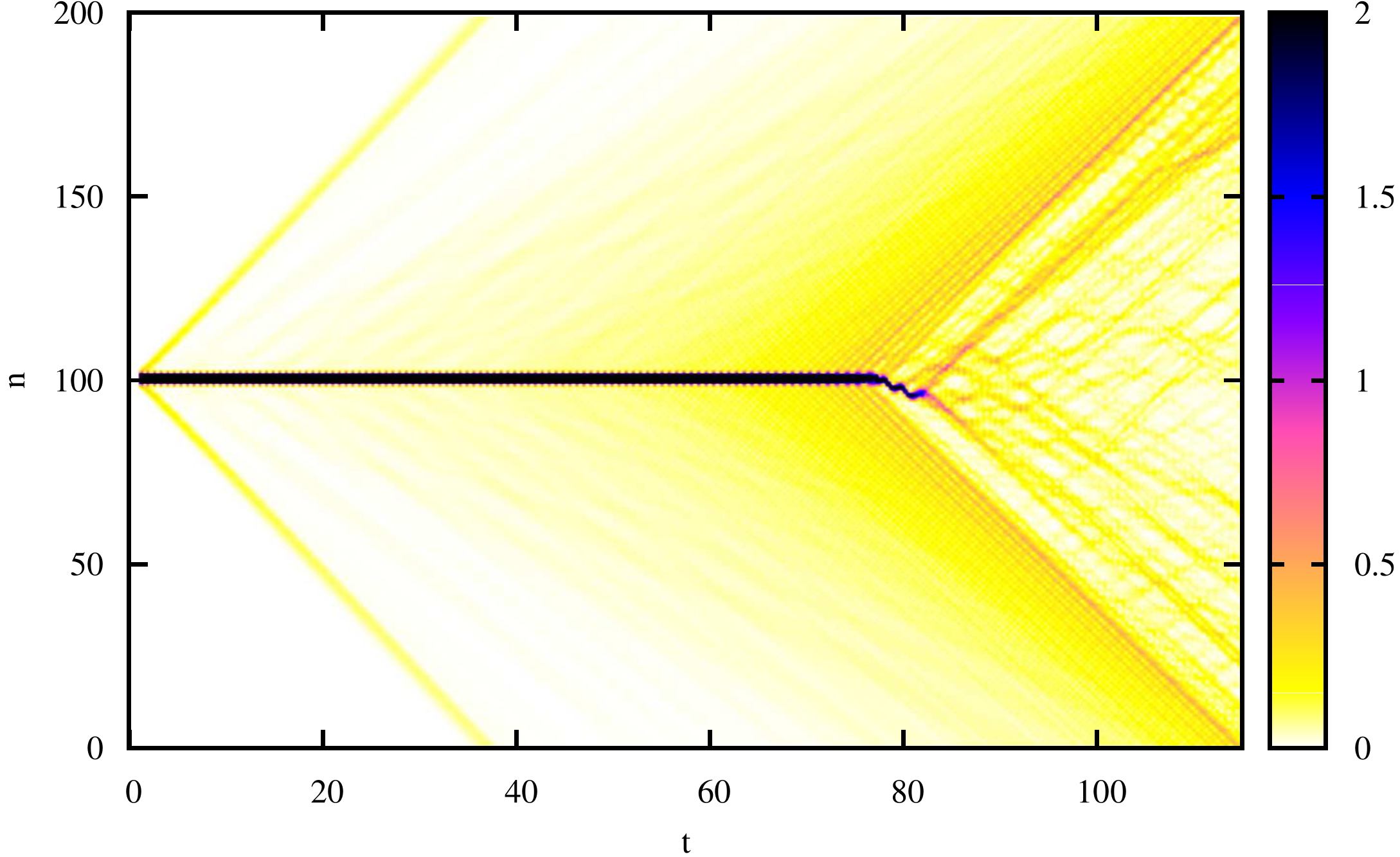}}%
  \caption{Soliton modes in the dynamic Fibonacci chain. Both
    modes radiate phonons.
    \label{fig:dfcmodes}}
\end{figure}

By varying the excitation strength and excitation shape function, we can
generate single solitons and breathers. As seen in Fig.~\ref{fig:dfcmodeskink},
solitons usually pass each other with only little interaction.  However it is
also found that solitons are not stable. They continuously radiate phonons,
therefore losing energy during propagation and slowing down. A soliton cannot
exist with energy under a certain minium threshold and will eventually decay.
Similar results hold for breathers. In Fig.~\ref{fig:dfcmodesbreather} a stationary
one is shown. After radiating phonons for some time, it disappears at $t=80$.

Note that the mathematical notion of soliton and breather is reserved for
localized modes that do not decay in time. So, strictly speaking our modes are
not ideal solitons/breathers.  However, compared to the characteristic time of
the system, both are stable and thus we adopt the naming.
Especially in the case of high amplitude modes or in periodic chains (e.g.\
the periodic $LS$-chain), solitons and breathers are stable over very long
times.

\section{Conclusion}\label{sec:conclusion}
The aim of our studies was an improved understanding of phasonic flips on an atomistic level
and of the consequences of the anharmonic potentials that are typical for quasicrystals and
other systems which provide a complex energy landscape.

Our main model system, the dynamic Fibonacci chain\cite{dfcpaper}, is constructed by providing
particles assembled as a part of the one-dimensional quasiperiodic Fibonacci chain
with an anharmonic pair potential of the form $V(x)=x^4-2x^2$ (double well potential). This system
and simplified systems derived from it (short chains, periodic $LS$ chain) were studied both
analytically and numerically using molecular dynamics simulations.

For the short ``chains'' consisting of three particles we
found analytical solutions of the equations of motion with
all particles oscillating in phase which can also be used to describe collective modes of motion in
longer chains. As one might expect, the oscillation frequencies of these modes are energy dependend because
of the anharmonic potential.

For studying the basic modes of motion which this potential induces
we used the periodic $LS$ chain at $T=0$. It turned out that the system does not only show the
usual phonons but also two different soliton modes in an energy regime higher than the energy barrier of the potential:
Breathers which are
a localized oscillatory mode and kink solitons which can be viewed as a propagating topological defect ($LS$
environments flip to $SL$ environments and vice versa), or a propagating phason flip.
These
modes are quite unstable and decay quickly. The kink solitons lose energy while propagating through the discrete
lattice. Breathers also radiate when propagating through the chain, but tend to get trapped at a certain particle
position; they can
be kept stable in a very defined and shielded environment.

We did not find any analytical description of the kink solitons, but we succeded in finding an analytical
approximate solution describing a breather.

Our studies of the dynamic Fibonacci chain in thermodynamical equilibrium showed that there
exists a lower threshold energy $\frac89$ for atomic flips (for both flip types: autonomous flips
and induced flips) which determines many properties of the system. E.g. it is important for transition of
the Fibonacci chain to a random tiling (our potential does not penalize $SS$ environments which are not
allowed in the Fibonacci chain!). We also found localized modes, which allow energy to be transported
through the chain. If the temperature gets higher, an increasing number of these modes reaches the
threshold energy, which leads to a large number of flips: The flip frequencies follow an Arrhenius law
for low temperatures.

In the dynamic Fibonacci chain, we observed the same modes of motion as in the $LS$ chain.
For observing solitary modes such as breathers and kinks, the interaction potential is more important than
the initial configuration. The same holds for thermodynamic properties: Specific heat increases above the
harmonic value solely because of the interaction potential. Thus it is not the phason degree of freedom but
the nonlinearity of the potential which causes the rise above the Dulong-Petit value.

The standard hydrodynamic theory of quasicrystals\cite{paphyd} predicts only diffusive phasons.
If one is leaving this long wavelength harmonic theory and proceeds to high excitation,
then propagating modes appear in our model. It is an interesting question, whether such
modes will be observable in highly excited quasicrystals.



\end{document}